\documentclass[%
 reprint,
superscriptaddress,
 amsmath,amssymb,
 aps,
]{revtex4-1}

\usepackage{amsmath}
\usepackage{mathtools}
\usepackage{physics}
\usepackage{lipsum}
\usepackage{dsfont}%remove if it creates conflict
\usepackage{nicematrix}%for better layout of some ugly matrices
\usepackage{bbold}
\usepackage{tikz}
\usepackage{graphicx}% Include figure files
\usepackage{dcolumn}% Align table columns on decimal point
\usepackage{bm}% bold math
\usepackage{placeins}
\usepackage{xcolor}
\usepackage{circuitikz}
\usepackage{bibentry}
\usepackage{ulem}
\usepackage[colorlinks,linkcolor=black]{hyperref}% 
\usepackage{orcidlink}

\allowdisplaybreaks

\newcommand{\ii}{\mathrm{i}}

\def \FigureSSHone
{
\begin{figure}[t]
    \centering
    \includegraphics[width=\linewidth]
    {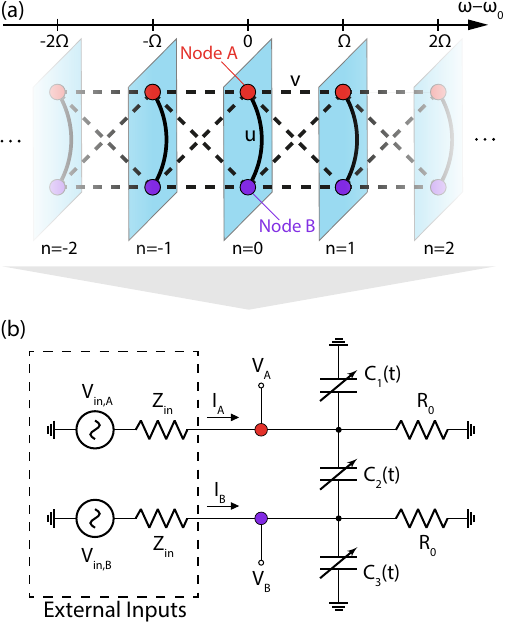}
    \caption{(a) Floquet SSH model in frequency space. The SSH lattice is mapped onto intra-site coupling $u$ (solid lines) connecting nodes $\mathrm{A}$ and $\mathrm{B}$, and inter-site couplings $v$ (dashed lines) between neighbouring frequency sites located at $n\Omega$. (b) Floquet SSH circuit. For each frequency, the two nodal voltages are $V_\mathrm{A}$ and $V_\mathrm{B}$; the input currents of the circuit are $I_\mathrm{A}$ and $I_\mathrm{B}$. Three capacitors $C_1(t)$, $C_2(t)$, and $C_3(t)$ constitute the time-varied capacitance matrix $\Gamma(t)$ in eq.~\eqref{diffequcirc}. } 
    \label{figsshfloqu1}
\end{figure}
}

\def \FigureSSHtwovtwo
{
\begin{figure*}[t]
    \centering
    \includegraphics[width=\linewidth]
    {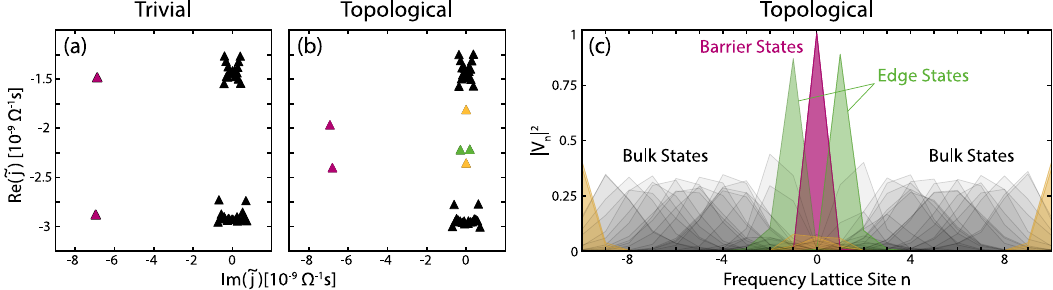}
    \caption{
    Spectrum and eigenstates of the measured normalized Floquet Laplacian $\tilde{\mathcal{J}}(\omega_0)_{nm}$ of the circuit for $\omega_0=0.05\Omega$. (a) Spectrum  of $\tilde{\mathcal{J}}(\omega_0)_{nm}$ for the topologically trivial case with $u=-0.75$ and $v=-0.25$. The barrier bound states (purple) are clearly separated from the gapped bulk states (black). (b) Spectrum of $\tilde{\mathcal{J}}(\omega_0)_{nm}$ for the topologically non-trivial case with $u=-0.25$ and $v=-0.75$. Additional mid-gap states (green, yellow) are observed. (c) Weight distribution of the eigenstates of $\tilde{\mathcal{J}}(\omega_0)_{nm}$ in the frequency lattice for the topological case. The mid-gap states marked green are found to be exponentially localized adjacent to the potential barrier at $n=0$, where the pair of bound barrier states (purple) resides. The mid-gap states marked yellow reside at the experimental frequency cutoff at $n=\pm10$, so their presence is an artefact of our measurement protocol. The remaining states (black) spread over the bulk of the frequency lattice.
    }
    \label{figsshfloqu2}
\end{figure*}
}

\def \FigureSSHthree
{
\begin{figure*}[t]
    \centering
    \includegraphics[width=\linewidth]
    {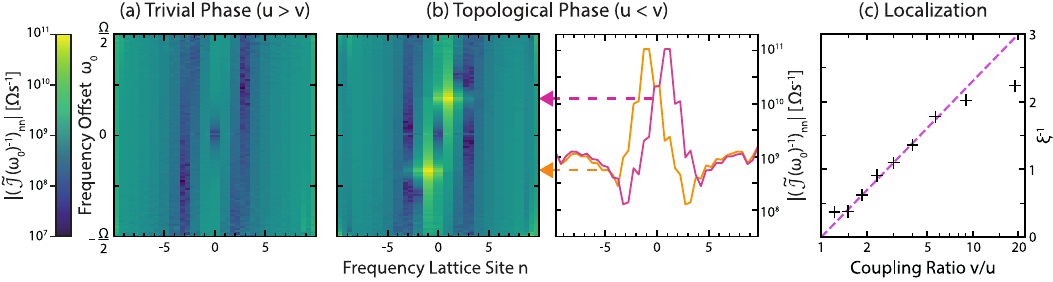}
    \caption{Measured frequency sweep over the Floquet Brillouin zone of the on-site impedances over the Floquet lattice. At each frequency lattice site the values for the sublattice circuit nodes A and B are shown. (a) and (b) depict the absolute values of the diagonal entries of the inverse normalized Floquet Laplacian $|(\tilde{\mathcal{J}}(\omega_0))^{-1}_{nn}|$ scaled logarithmically for the topologically trivial (a) and topological (b) phases with $u=-0.75,v=-0.25$ and $u=-0.25,v=-0.75$, respectively.  In the topological case, we observe resonance peaks that are exponentially localized at $n=\pm 1$. The right panel of (b) shows a horizontal cross section at the two $\omega_0$ of maximum impedance, indicated by the dashed arrows. In the logarithmic representation, we observe an exponential decay with the distance from the frequency barrier. (c) Fitted inverse localization length $\xi^{-1}$ for a set of measurements with different coupling ratios $v/u$. The observed resonance profiles (black markers) generally agree with the theoretical localization length of the SSH topological edge states (dashed line).
    }
    \label{figsshfloqu3}
\end{figure*}
}

\begin{document}

%\preprint{APS/123-QED}

%\title{Circuit Floquet SSH model}
\title{Topological Edge State Nucleation in Frequency Space\\ and its Realization with Floquet Electrical Circuits}

\author{Alexander Stegmaier\,\orcidlink{0000-0002-8864-5182}}
\email{alexander.stegmaier@uni-wuerzburg.de}
\affiliation{Institute for Theoretical Physics and Astrophysics, University of W\"urzburg, Am Hubland, D-97074 W\"urzburg, Germany}
\author{Alexander Fritzsche}
\affiliation{Institute for Theoretical Physics and Astrophysics, University of W\"urzburg, Am Hubland, D-97074 W\"urzburg, Germany}
\author{Riccardo Sorbello}
\affiliation{Institute for Theoretical Physics and Astrophysics, University of W\"urzburg, Am Hubland, D-97074 W\"urzburg, Germany}
\author{Martin Greiter}
\affiliation{Institute for Theoretical Physics and Astrophysics, University of W\"urzburg, Am Hubland, D-97074 W\"urzburg, Germany}
\author{Hauke Brand\, \orcidlink{0000-0003-4209-2663}}
\affiliation{Physikalisches Institut and R\"ontgen Research Center for Complex Material Systems, Universit\"at W\"urzburg, D-97074 W\"urzburg, Germany}
\author{Christine Barko}
\affiliation{Physikalisches Institut and R\"ontgen Research Center for Complex Material Systems, Universit\"at W\"urzburg, D-97074 W\"urzburg, Germany}
\author{Maximilian Hofer\,\orcidlink{0009-0005-3647-9353}}
\affiliation{Physikalisches Institut and R\"ontgen Research Center for Complex Material Systems, Universit\"at W\"urzburg, D-97074 W\"urzburg, Germany}
\author{Udo Schwingenschl\"ogl}
\affiliation{Physical Science and Engineering Division, King Abdullah University of Science and Technology (KAUST), Thuwal 23955, Saudi Arabia}
\author{Roderich Moessner}
\affiliation{Max Planck Institute for the Physics of Complex Systems, N\"othnitzer Straße 38, 01187 Dresden, Germany}
\affiliation{W\"urzburg-Dresden Cluster of Excellence ct.qmat, D-97074 W\"urzburg, Germany}

\author{Ching Hua Lee\,\orcidlink{0000-0003-0690-3238}}
\email{phylch@nus.edu.sg}
\affiliation{Department of Physics, National University of Singapore, Singapore, 117542}

\author{Alexander Szameit}
\affiliation{W\"urzburg-Dresden Cluster of Excellence ct.qmat, D-97074 W\"urzburg, Germany}
\affiliation{Institute of Physics, University of Rostock, Rostock, Germany}

\author{Andrea Al\`u}
\affiliation{Photonics Initiative, Advanced Science Research Center, City University of New York, New York, NY 10031, USA.}
\affiliation{Physics Program, Graduate Center, City University of New York, New York, NY 10016, USA.}

\author{Tobias Kie{\ss}ling}
\affiliation{Physikalisches Institut and R\"ontgen Research Center for Complex Material Systems, Universit\"at W\"urzburg, D-97074 W\"urzburg, Germany}
\affiliation{W\"urzburg-Dresden Cluster of Excellence ct.qmat, D-97074 W\"urzburg, Germany}

\author{Ronny Thomale\,\orcidlink{0000-0002-3979-8836}}
\email{ronny.thomale@uni-wuerzburg.de}
\affiliation{Institute for Theoretical Physics and Astrophysics, University of W\"urzburg, Am Hubland, D-97074 W\"urzburg, Germany}
\affiliation{W\"urzburg-Dresden Cluster of Excellence ct.qmat, D-97074 W\"urzburg, Germany}

\date{\today}

\begin{abstract}
We build Floquet-driven capactive circuit networks to realize topological states of matter in the frequency domain. We find the Floquet circuit network equations of motion to reveal a potential barrier which effectively acts as a boundary in frequency space.  By implementing a Su-Shrieffer-Heeger Floquet lattice model and measuring the associated circuit Laplacian and characteristic resonances, we demonstrate how topological edge modes can nucleate at such a frequency boundary.
\end{abstract}

\pacs{Valid PACS appear here}% PACS, the Physics and Astronomy
\maketitle

\paragraph*{Introduction.---}

Topological states of matter were initially explored within the realm of crystalline quantum materials.
There, the bulk-boundary correspondence provides a deep connection between bulk topology, which is typically resolved in momentum space, and protected modes at edges 
or defects~\cite{PhysRevD.13.3398,Su1979solitons,PhysRevB.25.2185,volovik2009universe,Hasan2010}. 
While the natural association with such a boundary would be some termination in real space, boundaries can likewise be realized in other spaces. For instance, a spatial dimension may be substituted by a time-like dimension which can then exhibit a boundary in frequency space~\cite{Dutt2022creating}.

The manifestation of topological matter in abstract spaces and artificial dimensions has proliferated into metamaterial platforms~\cite{Ozawa2019}. In systems such as photonic waveguides, cavities, and ring resonators, artificial dimensions and topological signatures thereof embody the tunability of optical degrees of freedom~\cite{Yuan18,Lustig21TopoPhot,Yang2022cavityrev,Ehrhardt2023} through utilizing mode dimensions~\cite{Lustig2019,Pirrucio2022cavities,Dutt2020,Yuan2021,Li2021,Cheng2023syndimband}, parameter modulations~\cite{Long23timebound,Liu2023blochsyndimwaveg}, or treating parameter space as an additional dimension~\cite{Nemirovsky2021}. Synthetic dimensions were also used in magneto-mechanical setups where topological pumping was observed~\cite{Grinberg2020} as well as atomic systems to study Bloch oscillations~\cite{Oliver2023}.

A fundamental resource providing non-spatial degrees of freedom is the time dimension, or its dual, frequency, where periodic driving is readily used to create lattices in frequency space. Floquet topological matter tends to feature concomitant phenomena such as artificial gauge fields~\cite{Rechtsman2013PFTI} and anomalous topological behaviour~\cite{Rudner2013anomal}. Compared to real space topological matter, frequency space realizations overcome certain limitations, e.g. naturally permitting non-local couplings through temporal engineering, and allow for higher-dimensional systems via multiple incommensurate driving frequencies~\cite{Martin2017,Nathan2022,Sridhar2024}.

In this Letter, 
we show how frequency space boundaries occur naturally in periodically driven electric circuit networks due to the unique structure of the underlying circuit network dynamics. 
The creation of such boundaries where topological frequency modes can nucleate and are amenable to observation is a crucial challenge of topological matter in synthetic dimensions.
This is particularly acute in periodically driven systems, as the driving implies an infinite, unbounded frequency lattice.
Previous proposals to overcome this challenge did not find an inherent solution, i.e. one which only involves the same degrees of freedom that form the topological state, but had to resort to additional setup components such as utilizing the  memory of the system~\cite{Baum18memory} and, as recently realized in optical ring resonators, the coupling to an auxiliary degree of freedom~\cite{Dutt2022creating}.

Through a combined theoretical and experimental effort we analyze the properties of periodically driven circuits, and demonstrate how they
provide the first instance of a natural emergence of frequency boundaries prone to nucleating topological modes.
Our work illustrates the versatility of driven circuits and their underlying formalism, which provides a framework for future realizations of novel topological phenomena set in frequency space.

\FigureSSHone

\paragraph*{Floquet Laplacian Formalism.---} 

Electric circuits are a versatile metamaterial platforms to realize topological phenomena~\cite{Lee2018circuits,Helbig2019,Ezawa2019,Olekhno2020,Wang2020_1,Helbig2020,Hofmann2020,Stegmaier2021,Zou2021,Yatsugi2022}. Typically, such experiments previously relied on static circuit setups. By employing active elements~\cite{Kotwal2021} such as analog multipliers~\cite{stegmaier2023thouless}, however, we propose a convenient way to induce periodic driving while preserving full experimental tunability. 
An $L$-$R$-$C$-circuit network is described by a system of second order differential equations 
\begin{align}
    \frac{\dd}{\dd t}  \vec{I}(t) = \frac{\dd^2}{\dd t^2}\Gamma \vec{V}(t) + \frac{\dd}{\dd t}\Sigma \vec{V}(t)+\Lambda \vec{V}(t).
    \label{diffequcirc}
\end{align}
$\vec{V}(t)$ denotes the voltages respective to ground at the circuit nodes and $\vec{I}(t)$ the external currents fed into the circuit nodes. The conductivity matrices $\Gamma$, $\Sigma$, and $\Lambda$ describe the resulting currents between the nodes through the capacitors ($\Gamma$), resistors ($\Sigma$), and inductors ($\Lambda$) of the circuit in response to the nodal voltages. In the Fourier domain, this equation takes the form
\begin{align}
    \vec{I}(\omega) = \ii \omega \Gamma \vec{V}(\omega) +\Sigma \vec{V}(\omega)+ \frac{1}{\ii \omega} \Lambda \vec{V}(\omega) = J(\omega) \vec{V}(\omega),
\end{align}
where the matrix $J(\omega)$ is called the circuit Laplacian~\cite{Lee2018circuits}.

In a Floquet circuit, elements can exhibit a periodic time dependence with a characteristic driving frequency $\Omega$. In the Fourier domain, this periodic modulation leads to mixing of the frequency components separated by multiples of $\Omega$, giving rise to an infinite lattice structure. The corresponding frequency lattice sites are located at $\omega_n = \omega_0 + n\Omega$, where $\omega_0$ is a frequency offset defined modulo $\Omega$. This is described by the infinite-dimensional Floquet Laplacian $\mathcal{J}(\omega_0)$ relation
\begin{align}
    \vec{I}_n(\omega_0) &=\sum_m \left(\mathrm{i}\omega_n\Gamma_{n-m}+\Sigma_{n-m}+\frac{\Lambda_{n-m}}{\mathrm{i}\omega_n}\right)\vec{V}_m(\omega_0)\nonumber\\
    &=\sum_m \mathcal{J}_{n m}(\omega_0) \vec{V}_m(\omega_0),
\end{align}
which is a matrix equation on the Floquet lattice~\cite{app}. It is analogous to the Floquet Schr\"odinger equation \mbox{$0 = \sum_m \left(H_{n-m} + (\varepsilon + n \Omega) \delta_{nm}\right) \ket{\psi_m(\varepsilon)}$} that describes periodic quantum systems ~\cite{Oka2019engineerQM}. The voltage vector $\vec{V}_m(\omega_0)$ takes the role of the state $\ket{\psi_m(\varepsilon)}$, $\omega_0$ replaces the quasi-energy $\varepsilon$, and there appears an additional source term, i.e., the external current $\vec{I}$. In contrast to a driven Hamiltonian, which always produces a linear potential in frequency space, the Floquet Laplacian can host terms with various frequency dependencies. Therefore, the theoretical framework of Floquet circuits is highly versatile, whose potential we advocate to realize a variety of topological models and phenomena.

\FigureSSHtwovtwo

\paragraph*{Floquet SSH circuit with frequency boundary.---}

We apply our framework to realize the topological edge states of the Su-Shrieffer-Heeger (SSH) model~\cite{Su1979solitons, Qiao2023degeneracy,li2023zak,Song2023AntiPTSSH} in frequency space using
a driven $C$-$R$-circuit.
The structure of the emergent Floquet lattice
is shown in Fig.~\ref{figsshfloqu1} (a). At each frequency lattice site $\omega_n$, there are two sublattice sites A and B, which exhibit an equal-frequency coupling $u$. Furthermore, there are couplings between 
neighboring frequency sites denoted by $v$. 

Our circuit realization of such a Floquet SSH model is sketched in Fig.~\ref{figsshfloqu1} (b). It has two circuit nodes that correspond to the sublattice sites in Fig.~\ref{figsshfloqu1} (a), carrying nodal voltages $\vec{V}(t)=(V_\mathrm{A}(t),V_\mathrm{B}(t))^\intercal$. At its core are three time-dependent capacitors, which are implemented based on four analog multipliers, so that the capacitances can be varied arbitrarily at high frequency through external control signals~\cite{app}.
The two function generators with added serial impedances $Z_\mathrm{in}$ provide input currents $\vec{I}(t)=(I_\mathrm{A}(t),I_\mathrm{B}(t))^\intercal$, and the resistors $R_0$ to ground realize a resistive on-site terms.

The capacitance matrix of our circuit $\Gamma(t)$ takes the form
\begin{equation}
\Gamma(t)=C_0\left[\left(\begin{array}{cc}
v\sin \Omega t & u+v\cos \Omega t \\ 
u+v\cos \Omega t & -v\sin \Omega t
\end{array} \right)+\alpha\mathds{1}\right]\,,
\label{gammassh}
\end{equation}
where $\alpha$ describes an offset that is the time-averaged capacitance between each node and ground. Choosing $\alpha>|u+v|$ ensures the dynamical stability of the circuit~\cite{app}.
Up to a Pauli rotation, Eq.~(\ref{gammassh}) corresponds to the conventional Bloch Hamiltonian representation of the SSH model. By analogy, the crystal momentum $k$ is replaced by time $t$, and the driving frequency $\Omega$ plays the role of the lattice constant. The coupling parameters $u,v\in \mathbb{R}$ can be tuned via the control signals at the analog multipliers ~\cite{app}. The resistive coupling matrix only consists of the two on-site terms, $\Sigma=R_0^{-1} \mathds{1}$, and the inductive term $\Lambda$ vanishes since no inductors are present in the circuit.

After Fourier transformation, the $C$-$R$-circuit differential equation reads
\begin{align}
 \frac{\vec{I}_n(\omega_0)}{\ii(\omega_0+n\Omega)}&=\sum_m\left[\Gamma_{n-m}+\frac{1}{\ii (\omega_0+n\Omega) R_\mathrm{0}}\delta_{nm} \mathds{1} \right]\vec{V}_m(\omega_0)\nonumber\\
&=\sum_m \tilde{\mathcal{J}}_{nm}\vec{V}_m(\omega_0),\label{eqbarrier}
\end{align}
where $\vec{I}_j(\omega_0)$ ($\vec{V}_j(\omega_0)$) is the $j$th component of the current (voltage) vector in the frequency lattice and $\tilde{\mathcal{J}}_{nm}$ is the normalized Floquet Laplacian ($\tilde{\mathcal{J}}_{nm}:=\mathcal{J}_{nm}/(\omega_0+n\Omega)$). Its first term $\Gamma_{n-m}$ corresponds to the $(n-m)$th Fourier component of Eq.~\eqref{gammassh} and forms a periodic hopping matrix coupling neighbouring frequency lattice sites (Fig.~\ref{figsshfloqu1} (a)). 

Importantly, for $\Omega \gg 1/(R_0 C_0),$ the second term $\delta_{nm}/(\ii (\omega_0+n\Omega)R_0)$ on the right hand side of Eq.~\eqref{eqbarrier} constitutes a potential barrier at $n=0$ for $\omega_0 \approx 0$. 
This potential barrier emerges naturally from the differential equation and effectively decouples the $0$th node from its neighbors, thereby introducing boundaries at $n = \pm 1$. This is crucial to the observation of boundary phenomena, since the underlying model without the resistive term is fully translationally invariant in frequency space.

\FigureSSHthree

\paragraph*{Observation of frequency space topological spectra.---}

We experimentally obtain the normalized admittance spectrum of the Floquet SSH circuit, which are the eigenvalues of $\tilde{\mathcal{J}}(\omega_0)$. 
This is done by measuring the voltage response to different input currents across the Floquet lattice. 
As the frequency lattice is infinite, we introduce a frequency cutoff for experimental measurements. 
For this, we use a sampling rate that is an integer multiple of the driving frequency, so that through aliasing the frequency domain of the measured signals becomes compactified. We choose the sampling frequency such that we obtain a frequency lattice of 21 sites. Further, in the circuit setup, an offset $\alpha=2$ (Eq.~(\ref{gammassh})) results in a shift along the real axis of $\tilde{j}$.

We distinguish the topologically trivial ($|u|>|v|$) from the non-trivial  ($|u|<|v|$) case depicted in Fig.~\ref{figsshfloqu2}(a) and (b), respectively. Most of the normalized admittance eigenvalues $\tilde{j}$ lie in the vicinity of the real axis, where they form a two-band gapped spectral structure (marked black).
The purple markers indicate two bound states of the barrier potential, that are separated from the remaining spectrum by an imaginary gap due to the high, imaginary-valued potential of the resistive term in Eq.~(\ref{eqbarrier}).
In the topological case (Fig.~\ref{figsshfloqu2}(b)), we observe mid-gap states in between the bulk bands.
By investigating the associated eigenvectors in Fig.~\ref{figsshfloqu2}(c), we find that the states marked yellow are localized at the artificial cutoff, and therefore not physical. The states marked green, however, localize at sites $n=\pm1$ adjacent to the potential barrier, which effectively forms a boundary for the left and right regions of the lattice. They are topological frequency mid-gap edge states of the SSH model.

\paragraph*{Topological Resonances at the Frequency Boundary.---}

In a second, independent measurement, we observe the signatures of the topological boundary states in frequency space by determining the resonances of the Floquet spectrum. Such resonances occur when the absolute value of an admittance eigenvalue of the Laplacian $|\tilde{j}_k|$ approaches $0$.
In the Floquet SSH model Eq.~(\ref{gammassh}) with no offset, $\alpha=0$, only the topological mid-gap states occur at approximately zero admittance, so we expect them to give rise to the resonances of the system.
In the case of an ideal potential barrier at zero frequency, the resonances would correspondingly lie at $\omega_0=0$. Outside of this limit, they move to higher $|\omega_0|$ for decreasing $R_0$ as the potential broadens. In the circuit implementation, setting $\alpha=0$ requires the introduction of an additional positive feedback that cancels constant on-site terms in the capacitance matrix~\cite{app}. 
Furthermore, we use capacitors $C_\mathrm{in} > C_0$ for the input impedances $Z_\mathrm{in}$ (see Fig.~\ref{figsshfloqu1} (b)) between the external voltage sources and the circuit nodes.
This compensates the tendency of the circuit to become dynamically unstable for $\alpha<|u+v|$.

To determine the resonances of the Floquet circuit, we measure the Floquet Laplacian over the whole range of the Floquet Brillouin zone 
$\omega_0\in [-\Omega/2,\Omega/2[$.
Experimentally, this is done by using a set of harmonically modulated Gaussian pulses 
as input signals, in order to probe the entire Floquet Brillouin zone simultaneously.  
From the resulting measurements we detect resonant modes by considering the diagonal elements of the inverse Floquet Laplacian, which are the impedances to ground at the different frequencies of the Floquet lattice:
\begin{align}
(\tilde{\mathcal{J}}(\omega_0))^{-1}_{nn}&=\left\langle n\right|\left(\sum_k \tilde{j}_k^{-1}\left|\psi_k\right\rangle\left\langle \psi_k\right|\right)\left|n\right\rangle\nonumber\\
&=\sum_k \tilde{j}_k^{-1} |\left\langle n| \psi_k \right\rangle|^2 \approx \tilde{j}_\mathrm{res}^{-1} |\left\langle n| \psi_\mathrm{res} \right\rangle|^2,\label{diagonal}
\end{align}
where $\left|\psi_k\right\rangle$ are the eigenvectors of the Floquet Laplacian.
The last approximation in Eq.~\eqref{diagonal} holds only if there is a 
resonant eigenvalue $|\tilde{j}_\mathrm{res}|\approx 0$. We see that, in this case, the diagonal elements of the inverse Floquet Laplacian are approximately proportional to the profile of the resonant eigenmode in the frequency lattice. 

The measured data is shown in Fig.~\ref{figsshfloqu3}. Since the admittance spectrum is never close to zero for the topologically trivial case (Fig.~\ref{figsshfloqu3} (a)), no resonances are expected to emerge, which is confirmed in our observations.
In the topological case (Fig.~\ref{figsshfloqu3} (b)), however, we find resonance peaks localized at the sites $n=\pm 1$, left and right of the barrier, which confirms that the topological boundary modes of the admittance spectrum give rise to resonances. Remarkably, {\it the scale of the dominant resonance frequency is determined by the driving frequency $\Omega$}, and not by the characteristic frequency scale suggested by dimensional analysis, $\omega_\text{ch} = (R_0 C_0)^{-1}$. This principle hence allows for the control of the resonance frequency via the external driving signal: In the limit $\Omega \gg \omega_\text{ch}$ and full dimerization $u=0, v=1$, the resonant mode will be a harmonic oscillation at frequency $\Omega$.

\paragraph*{Frequency localization length of the topological boundary modes.---}

The topological edge states of the SSH model decay exponentially with the distance from the boundary, where the localization length $\xi$ depends on the coupling ratio $v/u$. In the limit of $v/u\rightarrow \infty$, the topological modes are perfectly localized with $\xi=0$. For decreasing $v/u$, their localization length increases and the states become fully delocalized as the band gap closes for $v/u\rightarrow 1$.
To compare the localization length of the resonance profile to that of the SSH edge states, we repeat the described resonance measurement procedure for different coupling ratios.

We determine $\xi$ of the resonance profiles (compare Fig.~\ref{figsshfloqu3}(b) right panel) through a linear fit of the logarithm of the on-site impedances at the first three sites next to the barrier. 
Beyond $n=3$, the impedance peaks drop off so far that they are overshadowed by contributions of other non-resonant states.
The experimental results shown in Fig.~\ref{figsshfloqu3}(c) agree well with the theoretical prediction (denoted by the dashed purple line) for all but the lowest ($v/u = 11/9$) and the two highest ($v/u = 9$ and $19$) investigated coupling ratios. The localization lengths for high coupling ratios deviate from the theory as they are overshadowed by non-resonant states away from $|n|=1$ due to their strong localization. In the case of coupling ratios close to $1$, the band gap narrows, so that bulk modes at the band edges with eigenvalues $\tilde{j}$ approaching $0$ influence the impedance profile. These results confirm that the profile of the measured resonant mode over the Floquet lattice exhibits the characteristic shape of the SSH topological edge state.

\paragraph*{Conclusions.---}

We demonstrated how frequency space boundaries can naturally occur in periodically driven electric circuit networks. We showed that they are a result of the inhomogeneities intrinsic to the underlying Floquet Laplacian formalism, and therefore do not occur in Floquet Hamiltonians. We used this unique property of Floquet circuits to experimentally realize and observe the topological boundary states of an SSH chain in frequency space.

By exploiting the natural emergence of boundaries in the inherently infinite frequency space in active circuits, our work opens a pathway to study topological edge phenomena in synthetic spaces that are unattainable in most other experimental platforms due to the lack of accessible implementations of artificial boundaries. 
The unprecedented versatility of the underlying theoretical framework for Floquet circuits will enable the realization and study of a plethora of elusive topological phenomena in frequency space.

\paragraph*{Acknowledgments.---}

The work is funded by the Deutsche Forschungsgemeinschaft (DFG, German Research Foundation) through Project-ID 258499086 - SFB 1170, the W\"urzburg-Dresden Cluster of Excellence on Complexity and Topology in Quantum Matter -- \textit{ct.qmat} Project-ID 390858490 - EXC 2147, the Ministry of Education, Singapore (MOE award number: MOE-T2EP50222-0003), and through King Abdullah University of Science and Technology (KAUST).

\clearpage

\onecolumngrid

\appendix

\renewcommand{\thefigure}{A\arabic{figure}}

\setcounter{figure}{0}

\section{The Floquet Laplacian}

\subsection{Derivation for $L$-$R$-$C$-circuits}
The Floquet Laplacian $\mathcal{J}(\omega_0)$ is a generalization of the circuit Laplacian that can be used to describe the behavior of periodically driven electrical circuit networks.
Fundamentally, the Laplacian, or admittance matrix, describes the relation of external currents, collected in the current vector $\vec{I}(t)$, that are fed into the circuit nodes to the voltages between these nodes to ground, collected in the voltage vector $\vec{V}(t)$. The Laplacian is constructed by adding all internal voltages of the network flowing out of a respective node, and equating them to the external current flowing into that node, in accordance with Kirchhoff's current law \cite{Lee2018circuits}. For linear circuits, the internal currents flowing out of each node depend linearly on the nodal voltages, so an equation consisting of the external current vector, the nodal voltage vector, and a linear operator acting on the nodal voltages is obtained.
For a circuit network consisting of capacitors, resistors, and inductors, this takes the form of a linear differential equation
\begin{align}
\frac{\mathrm{d}}{\mathrm{d}t} \vec{I}(t) =  \frac{\mathrm{d}^2}{\mathrm{d}t^2} \Gamma \vec{V}(t) + \frac{\mathrm{d}}{\mathrm{d}t} \Sigma \vec{V}(t) + \Lambda \vec{V}(t).
\end{align}
The Laplacian is obtained by transforming this equation to the Fourier domain, giving
\begin{align}
\vec{I}(\omega) = \left[\mathrm{i}\omega \Gamma + \Sigma + \frac{1}{\mathrm{i}\omega} \Lambda \right] \vec{V}(\omega) = J(\omega) \vec{V}(\omega).
\end{align}
In a Floquet electrical circuit, every circuit element can in principle be time-dependent, i.e., time-dependent capacitors $\Gamma(t)$, resistors $\Sigma(t)$ or inductors $\Lambda(t)$ are all conceivable. One hurdle to writing down the correct equation of motion for the general case though is how the derivative (or integration) operators should be ordered with respect to the coefficient matrices, since these do not commute with them if time-dependent. One finds that this is ultimately a choice of convention since the equations can always be transformed into one or the other form, while proper care needs to be taken such that the equation of motion accurately reflects the underlying system. For example, an element intended to realize a Floquet capacitor may be implemented to behave according to $\vec{I}(t) = \frac{\mathrm{d}}{\mathrm{d}t} C(t) \vec{V}(t)$ or according to $\vec{I}(t) = C(t) \frac{\mathrm{d}}{\mathrm{d}t} \vec{V}(t)$, both of which could be re-expressed in either convention. For simplicity, we choose to put the differential operator in front of the coefficient matrix for the following calculations. For a derivative after the coefficient matrix, the only resulting difference is that in the derivation of Eq. (\ref{eq_app_floqlap}) the derivative ultimately becomes $\omega_0+m\Omega$ instead of $\omega_0+n\Omega$. 
In a periodically driven Floquet circuit, the coefficient matrix $M(t)$ is periodic with respect to the driving period $T$, so it can be decomposed into 
\begin{align}
M(t) = \sum_k M_k \, \mathrm{e}^{\mathrm{i}\Omega t},
\end{align}
with $\Omega=\frac{2\pi}{T}$ being the Floquet frequency.
Using this, we Fourier transform the equation of motion
\begin{align}
    \frac{\mathrm{d}}{\mathrm{d}t} \vec{I}(t) =  \frac{\mathrm{d}^2}{\mathrm{d}t^2} \Gamma(t) \vec{V}(t) + \frac{\mathrm{d}}{\mathrm{d}t} \Sigma(t) \vec{V}(t) + \Lambda(t) \vec{V}(t)
\end{align}
to obtain
\begin{align}
    \vec{I}(\omega) = \sum_k \left[\mathrm{i} \omega \Gamma_k + \Sigma_k + \frac{1}{\mathrm{i}\omega} \Lambda_k \right] \vec{V}(\omega - k\Omega).
\end{align}
We re-write the frequency $\omega$ as $\omega = \omega_0 + n\Omega$,\; $\omega_0 \in [0,\Omega[$. Furthermore, we introduce the notation $\vec{I}_n(\omega_0) = \vec{I}(\omega_0+n\Omega)$, $\vec{V}_n(\omega_0) = \vec{V}(\omega+n\Omega)$ and substitute $k=m-n$. Doing this, we can re-cast the problem in the form of an (infinite-dimensional) matrix-vector equation
\begin{align}
    \vec{I}_n(\omega_0) = \sum_m \left[\mathrm{i} (\omega_0+n\Omega) \Gamma_{m-n} + \Sigma_{m-n} + \frac{1}{\mathrm{i}(\omega_0+n\Omega)} \Lambda_{m-n} \right] \vec{V}_m(\omega_0)= \sum_m \mathcal{J}_{nm}(\omega_0) \vec{V}_m(\omega_0).\label{eq_app_floqlap}
\end{align}
$\mathcal{J}_{nm}(\omega_0)$ then are entries of the Floquet Laplacian, a tensor that connects the vectors of currents $\{\vec{I}_n(\omega_0)\}$ and voltages $\{\vec{V}_n(\omega_0)\}$ over the Floquet lattice that is spanned by the set of frequencies $\{\omega_0 + n\Omega, \, n\in\mathds{Z}\}$.
\section{Circuit Laplacian of the Floquet SSH Model}
In this section we derive the form of the Floquet Laplacian of the Su-Schrieffer-Heeger (SSH) lattice in frequency space.  We start from the SSH-Bloch-Hamiltonian matrix 
\begin{equation}
H_\mathrm{SSH}=\left(\begin{array}{cc}
0 & u+v\mathrm{e}^{-\ii k} \\ 
u+v\mathrm{e}^{\ii k} & 0
\end{array} \right)
\end{equation}
where $u$ and $v$ denote the intra and inter unit cell coupling, respectively, and $k$ is the quasi-momentum.
We then apply the unitary transformation $T=\dfrac{1}{\sqrt{2}}\left(\begin{array}{cc}
\ii & 1 \\ 
1 & \ii
\end{array} \right)$ to transform the Bloch-Hamiltonian into a form that is more feasible for an experimental implementation in electric circuits
\begin{equation}
H'_\mathrm{SSH}=\left(\begin{array}{cc}
v\sin k & u+v\cos k \\ 
u+v\cos k & -v\sin k
\end{array} \right) .
\end{equation}
This coupling matrix is then realized in a $C$-$R$-circuit where the quasi-momentum is replaced by the time $t$ to realize an SSH chain in frequency space. The dynamics of the system are described by 
\begin{equation}
\frac{\mathrm{d^2}}{\mathrm{d}t^2}C_0\Gamma (t) \vec{V}(t)+\frac{\mathrm{d}}{\mathrm{d}t}\frac{1}{R_0}\vec{V}(t)=\frac{\mathrm{d}}{\mathrm{d}t}\vec{I}(t)
\label{diffequapp}
\end{equation}
where 
\begin{equation}
\Gamma (t)=\left(\begin{array}{cc}
v\sin \Omega t & u+v\cos \Omega t \\ 
u+v\cos \Omega t & -v\sin \Omega t
\end{array} \right) , \quad \vec{V}(t)=\left(\begin{array}{c}
V_\mathrm{A}(t) \\ 
V_\mathrm{B}(t)
\end{array} \right) , \quad \mathrm{and} \quad\vec{I}(t)=\left(\begin{array}{c}
I_\mathrm{A}(t) \\ 
I_\mathrm{B}(t)
\end{array} \right).
\end{equation}
Here, $\Gamma(t)$ denotes a set of time-dependent capacitors and implements the SSH model. The capacitors are modulated with driving frequency $\Omega$ such that $\Gamma(t)=\Gamma(t+2\pi/\Omega)$. $\vec{V}(t)$ and $\vec{I}(t)$ are the voltage and current vectors at the electric nodes A and B that make up the unit cell. We then decompose the matrix $\Gamma (t)$ into its Fourier components $\Gamma (t)=\sum_{l=-1}^{1}\Gamma_l\mathrm{e}^{\ii l\Omega t}$ with
\begin{equation}
\Gamma_{-1}=\left(\begin{array}{cc}
\ii v/2 & v/2 \\ 
v/2 & -\ii v/2
\end{array} \right) , \quad \Gamma_0=\left(\begin{array}{cc}
0 & u \\ 
u & 0
\end{array} \right) , \quad\mathrm{and} \quad \Gamma_{1}=\left(\begin{array}{cc}
-\ii v/2 & v/2 \\ 
v/2 & \ii v/2
\end{array} \right).
\end{equation}
We further Fourier transform the voltage and current as well as the differential equation (\ref{diffequapp}) and obtain
\begin{equation}
\frac{1}{2\pi}\iint \mathrm{d}t\mathrm{d}\nu\vec{V}(\nu)\left\lbrace\mathrm{e}^{-\ii\omega t}C_0\frac{\mathrm{d}^2}{\mathrm{d}t^2}\sum_l\Gamma_l\mathrm{e}^{\ii l\Omega t}\mathrm{e}^{\ii\nu t}+\frac{1}{R_0}\frac{\mathrm{d}}{\mathrm{d}t}\mathrm{e}^{\ii\nu t}\right\rbrace=\frac{1}{2\pi}\iint \mathrm{d}t\mathrm{d}\nu \mathrm{e}^{-\ii\omega t}\frac{\mathrm{d}}{\mathrm{d}t}\mathrm{e}^{\ii\nu t}\vec{I}(\nu)
\end{equation}
which then becomes
\begin{equation}
\iint \mathrm{d}t\mathrm{d}\nu\vec{V}(\nu)\left\lbrace\sum_l-(-\Omega l+\nu)^2\mathrm{e}^{\ii t(l\Omega +\nu-\omega)}C_0\Gamma_l+\ii\nu\frac{1}{R_0}\mathrm{e}^{\ii t(\nu-\omega)}\right\rbrace=\iint \mathrm{d}t\mathrm{d}\nu \ii \nu\mathrm{e}^{\ii t(\nu-\omega)}\vec{I}(\nu)
\end{equation}
\begin{equation}
\int\mathrm{d}\nu\vec{V}(\nu)\left\lbrace\sum_l-\delta(\omega-l\Omega-\nu)(-l\Omega+\nu)^2C_0\Gamma_l+\frac{1}{R_0}\delta(\omega-\nu)\ii\nu\right\rbrace=\int\mathrm{d}\nu \ii\nu\delta(\omega-\nu)\vec{I}(\nu)
\end{equation}
\begin{equation}
\sum_l-C_0\Gamma_l\omega^2\vec{V}(\omega-l\Omega)+\frac{1}{R_0}\ii\omega\vec{V}(\omega)=\ii\omega\vec{I}(\omega).
\end{equation}
We further rewrite $\omega=\omega_0+n\Omega$ where $\omega_0$ lies between $-\Omega/2$ and $\Omega/2$ such that
\begin{equation}
-\sum_l C_0\Gamma_l(\omega_0+n\Omega)^2\vec{V}(\omega_0+(n-l)\Omega)+\frac{1}{R_0} \ii (\omega_0+n\Omega)\vec{V}(\omega_0+n\Omega)=\ii(\omega_0+n\Omega)\vec{I}(\omega_0+n\Omega).
\end{equation}
This equation is then divided by $(\omega_0+n\Omega)^2$ with $m=n-l$ as well as the definition $\vec{V}(\omega_0+j\Omega)=\vec{V}_j(\omega_0)$ (similarly defined for the current $\vec{I}$). We then obtain the infinite dimensional matrix equation
\begin{equation}
\sum_m\left[C_0\Gamma_{n-m}+\frac{1}{\ii(\omega_0+n\Omega)R_0}\delta_{nm}\right]\vec{V}_m(\omega_0)=\sum_m \tilde{\mathcal{J}}_{nm}\vec{V}_m(\omega_0)=\frac{1}{\ii(\omega_0+n\Omega)}\vec{I}_n(\omega_0).
\end{equation}
In this equation, $\vec{I}_j(\omega_0)$ and $\vec{V}_j(\omega_0)$ are components of an infinite vector and $\tilde{\mathcal{J}}_{nm}$ is the normalized circuit Floquet Laplacian. In the experiment, a finite section of this matrix is measured. 
The Laplacian consists of two terms. $C_0\Gamma_{n-m}$ is the $(n-m)$th component of the SSH hopping matrix $C_0\Gamma(t)$ and $1/(\ii(\omega_0+n\Omega)R_0)$ constitutes a potential barrier at $n=0$ if $\omega_0\ll \Omega$ that effectively acts as a boundary of the SSH chain.
\begin{figure}
    \centering
    \includegraphics[width=\textwidth]{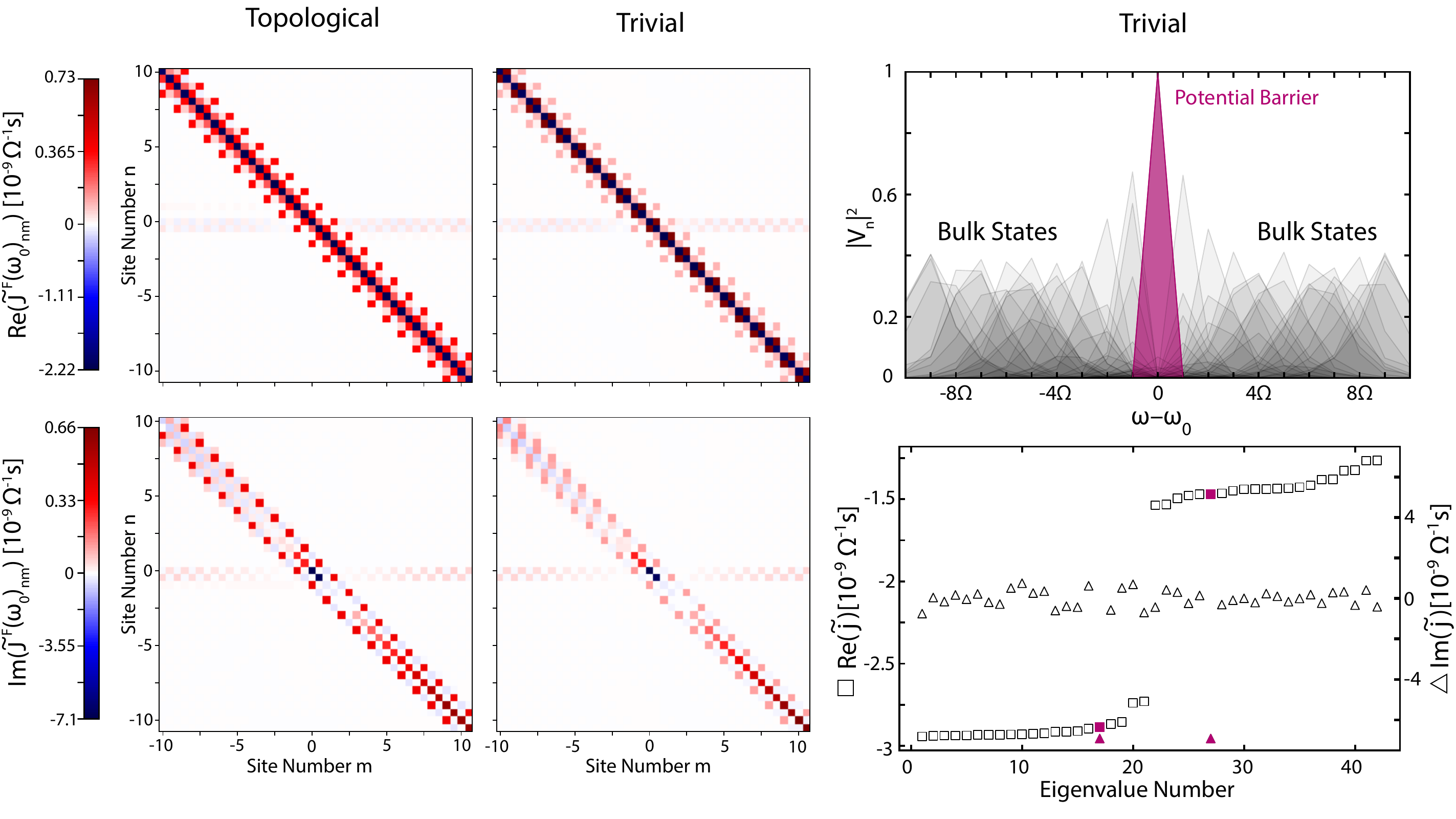}
    \caption{Matrix plot of the measured $\tilde{\mathcal{J}}_{nm}$ for the topologically non-trivial (left) and trivial (right) case. The real part is shown in the top panels and the imaginary part in the bottom panels. Here, the extremal values of the scales represent the maximum and minimum values of $\tilde{\mathcal{J}}_{nm}$. Note the different scales for the real and imaginary parts. The bandstructure and eigenstate localization for the topologically trivial case are shown on the right. The bulk bands (black rectangles) are gapped and do not have a significant imaginary part (black triangles). The potential barrier, however, is characterized by a significant negative imaginary part (purple triangles). In contrast to the topological case (main text Fig. \ref{figsshfloqu2}) there are no mid-gap states and no edge states that are localized at the potential barrier.}    \label{figsshfloquapp}
\end{figure}

\section{Resonances, dynamical eigenmodes, and stability}

\subsection{Theoretical eigenmodes of the system}
The behavior of the dynamical eigenmodes, or resonant states, of a Floquet circuit are given by its Floquet multipliers $\lambda = \mathrm{e}^{- \mathrm{i} \varepsilon T}$. We can evaluate these using two different approaches: One is to calculate the Floquet operator $U$, describing the evolution of the system over one driving period, and the other is to find the roots of the admittance spectrum as $\omega_0$ is varied. The first approach provides a more straightforward numerical recipe whereas the second one allows us to draw connections between the resonant states and the spectrum and eigenvectors of the Laplacian.

In order to calculate the Floquet operator $U$, we bring the homogeneous Floquet-SSH-circuit equation of motion
\begin{align}
    \vec{0} = \frac{\mathrm{d}}{\mathrm{d}t} \Gamma(t) \vec{V}(t) + \Sigma \vec{V}(t) = \frac{\mathrm{d}}{\mathrm{d}t} 
    C_0 \begin{pmatrix} v\sin \Omega t & u+v\cos \Omega t \\ u+v\cos \Omega t & -v\sin \Omega t  \end{pmatrix} \vec{V}(t) + \frac{1}{R_0} \vec{V}(t)
\end{align}
into the form
\begin{align}
 \frac{\mathrm{d}}{\mathrm{d}t} \vec{V}(t) = - \Gamma^{-1}(t) \left(\dot{\Gamma}(t) + \Sigma\right) \vec{V}(t) = A(t) \vec{V}(t).
\end{align}
We can now evaluate the Floquet operator through the time evolution operator over one period $T=\frac{2\pi}{\Omega}$
\begin{align}
    U = \mathcal{T} \exp{\int_0^T A(t) \mathrm{d}t} = \lim_{N\to\infty} \prod_{n=0}^{N-1} \exp{A(n T / N)\, T / N}.
\end{align}
%\paragraph*{With offset}
%PUT TEXT ABOUT STABILITY WITH OFFSET HERE
%EIGVALS OF GAMMA POSITIVE SO NO IMMEDIATE INSTABILITY EXPECTED EXPECTED, RESULTS FOR FLOQUET MULTIPLIERS
%\paragraph*{Without offset}
We now insert the circuit parameters used in the resonance measurement, $\Omega = 2\pi \cdot 10^4 \,\mathrm{Hz},\; C_0 = 10^{-9}\,\mathrm{F},\; R_0= 10^5\,\mathrm{\Omega}$, $\alpha=0$, and \mbox{$u=0.25,\; v=0.75$} for the topological case or \mbox{$u=0.75,\; v=0.25$} for the trivial case. The result for the topological case is that the Floquet multipliers have an absolute value of one, meaning stable oscillations, and the quasi eigenfrequencies $\varepsilon = \frac{\arg{(\lambda)}}{2\pi}\Omega$ are $\pm 0.0503 \, \Omega$. In the trivial case, the absolute values of the Floquet multipliers are $|\lambda_1| = |1/\lambda_2| = 3.817$ and the quasi-eigenfrequencies are zero.
If a stabilizing offset $\alpha > |u+v|$ is added to the capacitance matrix, the eigenvalues of $\Gamma(t)$ are positive at all times, so that the circuit will be dissipative over the entire Floquet period. Barring parametric resonance, which can be excluded since we choose the driving frequency to be much larger than the natural scale of resonance $\Omega \gg (R_0 C_0)^{-1}$, the Floquet multipliers must then have absolute values smaller than one.

\begin{figure}
    \centering
    \includegraphics[width=\linewidth]{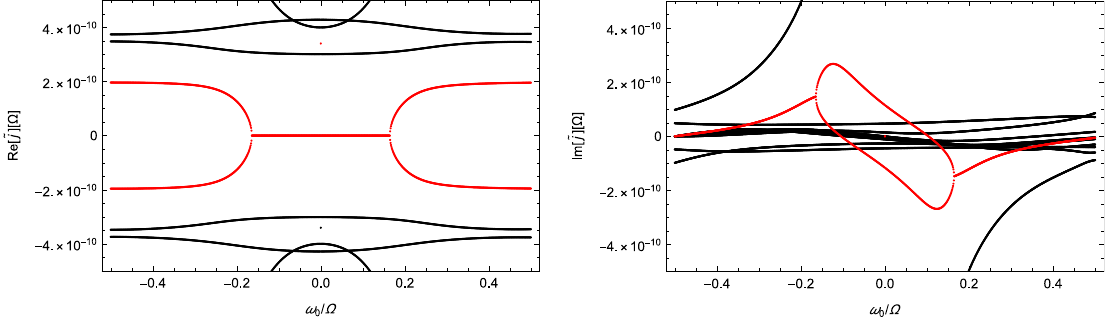}
    \caption{Real (left) and imaginary (right) part of the eigenvalues of the normalized Floquet Laplacian over $\omega_0$. Marked in red are the two states that correspond to the topological mid-gap states.}
    \label{res_eigvals}
\end{figure}

Next, let us consider the admittance picture, i.e., the eigenvalues of the Laplacian, to better understand the relationship between the admittance states of the model and its resonances. Figure \ref{res_eigvals} shows the eigenvalue spectrum of $\tilde{\mathcal{J}}(\omega_0)$ for $\omega_0 \in [-\Omega/2, \Omega/2[$.
Marked in red are the two bands from which the mid-gap states emerge in the vicinity of $\omega_0 \approx 0$.
We observe that these eigenvalues are first split in their real part, for relatively large values of $\omega_0$, i.e., in the regime where the potential barrier is absent from the lattice and these states are delocalized. As $\omega_0$ approaches zero, however, they approach each other and then meet in a branch point, where they split in the imaginary part while the real part remains zero. The imaginary contribution can be associated with the resistive term in the Laplacian $\frac{1}{\mathrm{i}\omega_n R_0}$ that forms the potential barrier. While this term is largest at Floquet site $n=0$, it is still non-negligible at $n = \pm 1$, which causes the imaginary shift in the topological states localized there. Further approaching $\omega_0=0$, the eigenvalues then truly cross zero and become resonant states. Accordingly, the resonant modes emerging from this will show the shape of the exponentially localized topological boundary modes at $n = \pm 1$ in the Floquet lattice.

\subsection{Instability of realistic circuits for $\alpha=0$}
Despite the theoretically stable oscillations, we find that in a realistic circuit setup unstable behavior emerges. This can be explained as follows. Consider a small higher order perturbation $\epsilon \frac{\mathrm{d}^{k+1}}{\mathrm{d}t^{k+1}} \vec{V}(t),\; 0 < \epsilon \ll 1$, to be added to the equation of motion. Then, we investigate the short-time dynamics for some $t\approx t_0$, so that $\Gamma(t)\approx \Gamma(t_0)$. The eigenvalues of $\Gamma(t_0)$ are $C_0 \gamma_\pm = \pm C_0 \sqrt{u^2 + v^2 + 2 u v \cos{\Omega t_0}}$. If we choose the eigenvector associated with the negative eigenvalue $\vec{v}_-(t_0)$ and choose the ansatz $\vec{V}(t) = \vec{v}_-(t_0) \mathrm{e}^{\lambda t}$, we obtain the reduced equation of motion
\begin{align}
    0&=\epsilon \frac{\mathrm{d}^{k+1}}{\mathrm{d}t^{k+1}} \vec{v}_-(t_0) \mathrm{e}^{\lambda t} + C_0 \gamma_- \frac{\mathrm{d}}{\mathrm{d}t} \vec{v}_-(t_0) \mathrm{e}^{\lambda t} + \frac{1}{R_0} \vec{v}_-(t_0) \mathrm{e}^{\lambda t} 
    = (\epsilon \lambda^{k+1} + C_0 \lambda \gamma_- +\frac{1}{R_0})\vec{v}_-(t_0) \mathrm{e}^{\lambda t}\\
    \iff 0 &= \lambda^{k+1} + \frac{C_0 \lambda}{\epsilon} \gamma_- +\frac{1}{\epsilon R_0} \approx \lambda^{k+1} + \frac{C_0\lambda}{\epsilon} \gamma_-\\
    \iff \lambda &\approx \sqrt[k]{-\frac{C_0 \gamma_- }{\epsilon}}.
\end{align}
So, since $\epsilon \ll 1$ and $\gamma_-<0$, $\lambda$ is real-valued and large. This implies that a rapidly diverging dynamical eigenmode exists, so $\vec{V}(t)$ will be highly unstable. This divergent eigenmode is a new eigenmode introduced due to the increased order of the differential equation, such that in the impedance spectrum the ideal theoretical eigenmodes are still present, just in the dynamical behavior they are overshadowed by the diverging modes.
Here again, an offset $\alpha \geq 2(u+v)$ will stabilize the circuit dynamics since it shifts the eigenvalues of $\Gamma(t)$ to be non-negative at all times.

\section*{Measuring the Floquet Laplacian}
\subsection{General measurement protocol}
The Floquet Laplacian $\mathcal{J}(\omega_0)$ can be measured by treating the sites $\omega_n$ of the Floquet lattice like regular lattice sites. The values of the currents and voltages at these sites are their corresponding frequency components at the frequencies $\omega_n$. Since the number of different $\omega_n$ is in principle infinite, some kind of truncation needs to be applied to obtain a finite lattice. For now, we simply choose to discard any frequency components beyond some arbitrary cutoff frequency $\omega_\mathrm{min} =\omega_0 + n_\text{min}\Omega \leq \omega < \omega_{\text{max}} = \omega_0 + n_\text{max}\Omega, \quad n_{\text{max}}-n_{\text{min}} = k$. We obtain a set of Floquet lattice sites $\{\omega_n\} = \{\omega_0 + n_\text{min}\Omega, \omega_0 + (n_\text{min}+1) \Omega, \ldots, \omega_0 + (n_\text{max}-1)\Omega\}$.
Next, consider the vectors $\vec{\mathcal{I}}$ and $\vec{\mathcal{V}}$, that combine all current and voltage components $\vec{I}(\omega_n) =: \vec{I}_n(\omega_0)$, $\vec{V}(\omega_n) =: \vec{V}_n(\omega_0)$ spanning the measurement range, 
\begin{align}
    \vec{\mathcal{I}}(\omega_0) &:= \left(\vec{I}_{n_\text{min}}(\omega_0)^\top, \ldots, \vec{I}_{n_\text{max}}(\omega_0)^\top \right)^\top\\
    \vec{\mathcal{V}}(\omega_0) &:= \left(\vec{V}_{n_\text{min}}(\omega_0)^\top, \ldots, \vec{V}_{n_\text{max}}(\omega_0)^\top \right)^\top.
\end{align}
Now, we perform a sequence of measurements where we apply $N k$ linearly independent currents $\vec{\mathcal{I}}_i$, where $N$ is the number of sites in the circuit, and obtain the corresponding responses $\vec{\mathcal{V}}_i$. Then, the Floquet Laplacian can be reconstructed via
\begin{align}
    \mathcal{J}(\omega_0) = \left(\vec{\mathcal{I}}_1(\omega_0), \vec{\mathcal{I}}_2(\omega_0), \ldots, \vec{\mathcal{I}}_{N k}(\omega_0)\right) \left(\vec{\mathcal{V}}_1(\omega_0), \vec{\mathcal{V}}_2(\omega_0), \ldots, \vec{\mathcal{V}}_{N k}(\omega_0)\right)^{-1}.
\end{align}

For measurements with a single offset frequency $\omega_0$, an appropriate measurement protocol is to use harmonic excitations at the considered frequencies of the Floquet lattice between $\omega_\text{min}$ and $\omega_\text{max}$ at one node at a time, and zero input at the other nodes. Repeating this for each circuit node obtains the desired $N k$ linearly independent measurements.

%Then we combine the corresponding voltages $\{\vec{I}(\omega_n)=\vec{I}_n(\omega_0)\}$ and the vectors $\vec{V}^F(\omega_0)$, and correspondingly for the input currents $\vec{I}^F(\omega_0)$. The components of these vectors account for both the actual nodes of the circuit, that take on the role of a sub-lattice to the Floquet lattice, and the Floquet lattice sites.
%
%To obtain the components of these vectors in a measurement, consider the steady-state response of a Floquet circuit to some applied input signal. One measures the voltages and input currents of the circuit at the frequencies $(\omega_{-k},\omega_{-k+1},\ldots ,\omega_{k-1})$ and then constructs the Floquet-Laplacian from $2 k$ measurements with linearly independent input signals as
%\begin{align}
% \tilde{J}(\omega_0)=\left(\vec{I}_1, \vec{I}_2, \ldots \vec{I}_{2k}\right) \left(\vec{V}_1, \vec{V}_2, \ldots \vec{V}_{2k}\right)^{-1}.
%\end{align}

\subsection{Broadband excitation}
Instead of measuring the Floquet Laplacian for only one steady-state frequency, one can also consider broader excitations, for example a set of modulated Gaussian pulses of the form
\begin{align}
    A_0 \exp\left(- (t-t_0/\Delta)^2\right) \sin\left(k\Omega t+\varphi\right).
\end{align} 
This way, by selecting the corresponding frequency contributions from the measured data, the Floquet Laplacian $\mathcal{J}$ can be obtained for a whole range of $\omega_0$ in a single measurement series.

\subsection{Aliasing as a form of Floquet lattice regularization}
Previously, we considered that the Floquet lattice is reduced to a finite number of sites by introducing a frequency cutoff and discarding everything beyond. This method comes with a downside in that it creates a non-physical boundary on the frequency lattice that may cause undesired artifacts in, for example, calculated eigenvectors, especially if boundary states are of interest. Another method of reducing the infinite Floquet lattice to finite size is aliasing, where a finite sampling frequency $k\Omega$ is set such that frequencies separated by $k\Omega$ become indistinguishable, essentially wrapping the infinite lattice around a ring of $k$ sites. Formally, this means that in place of $\vec{I}_n(\omega_0)$ or $\vec{V}_n(\omega_0)$ we effectively measure
\begin{align}
\sum_{c=-\infty}^\infty \vec{I}_{n+ck}(\omega_0) \quad \text{and} \quad \sum_{c=-\infty}^\infty \vec{V}_{n+ck}.
\end{align}
A downside of this measurement procedure is that the result is no longer unique: If all input signals are shifted by the aliasing frequency, the resulting vector of input signals will look the same but the vector of responses may be different since the circuit can behave differently at those frequencies, yielding a different Laplacian. This means, that the range of excitation frequencies must be well-defined and should lie within the range of the aliasing frequency to obtain consistent results. At the point of identification, inhomogeneities in the model can still arise as exemplified by the additional artificial states in the gap observed in the main text, Fig.~2~(b,c).

\section{The Floquet SSH Circuit}

%\subsection{Circuit setup}

\begin{figure}[h]
\centering
\includegraphics[width=1.0\linewidth]{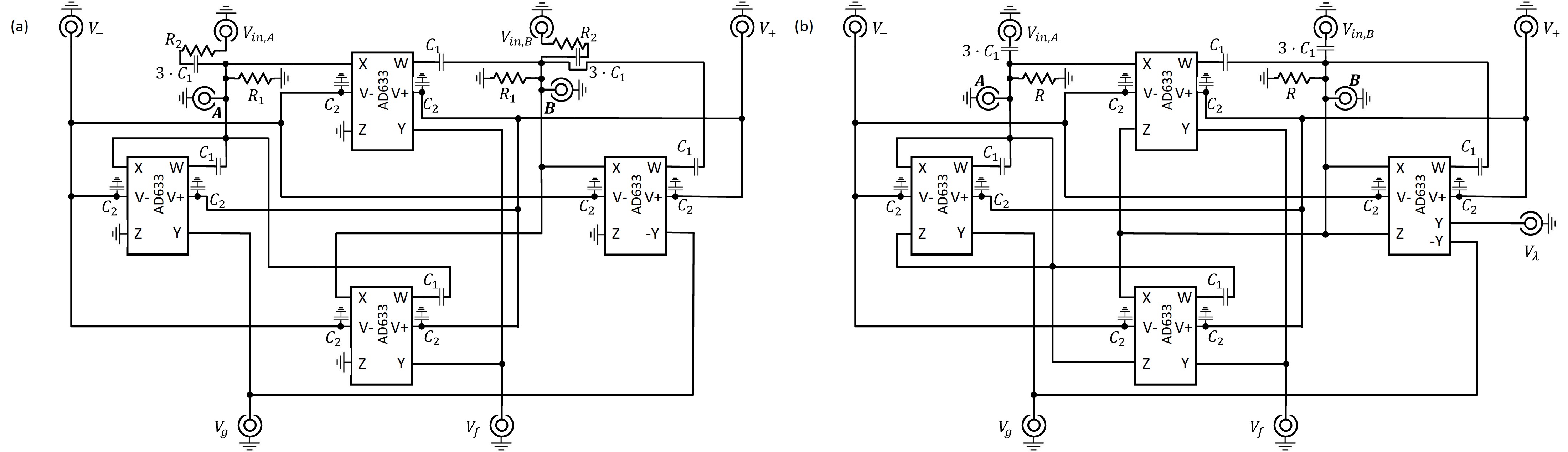}
\caption{
Detailed circuit implementation of the used variants of the Floquet SSH model.
Four analog multipliers build the core of both variants of these elements.
\textbf{Variant (a):}
The voltages $V_{\mathrm{in},A}$ and $V_{\mathrm{in},B}$ are fed into the circuit via $500~\Omega$ resistors, over which the current into the circuit is measured, in series with $3.3 \cdot C_1 = 3.3~\text{nF}$ capacitors for DC-decoupling.
The output W of each multiplier is connected to a capacitor of $C_1 = 1~\text{nF}$.
Node $A$ is coupled to the X input of the upper and the left multiplier, to the capacitors of the left and the lower multiplier and to a resistor $R = 110~\text{k}\Omega$ to ground.
Node $B$ is coupled to the X input of the right and lower multiplier, to the capacitors of the upper and right multiplier at their respective output W, as well as to another resistor $R = 50~\text{k}\Omega$ to ground.
The capacitors $C_2 = 1~\mu \text{F}$ to ground constitute low pass filters that shield the multipliers from picked up high frequency signals on the lines for the supply voltages.
The signal $V_g$ is fed directly into the Y input of the left and the negated Y input of the right multiplier, whereas the signal $V_f$ is fed into the Y input of the upper and the lower multiplier.
\textbf{Variant (b):}
The voltages $V_{\mathrm{in},A}$ and $V_{\mathrm{in},B}$ are fed into the circuit via $3.3 \cdot C_1 = 3.3~\text{nF}$ capacitors, over which the current into the circuit is measured.
The output W of each multiplier is connected to a capacitor of $C_1 = 1~\text{nF}$.
Node $A$ is coupled to the X input of the upper and the left multiplier, to the capacitors of the left and the lower multiplier, and to a resistor $R = 110~\text{k}\Omega$ to ground.
Additionally, node $A$ is coupled to the Z inputs of the lower and the left multiplier.
Node $B$ is coupled to the X input of the right and the lower multiplier, to the capacitors of the upper and the right multiplier at their respective output W, to another resistor $R = 110~\text{k}\Omega$ to ground, as well as to the Z inputs of the upper and the right multiplier.
The capacitors $C_2 = 1~\mu \text{F}$ to ground constitute low pass filters that shield the multipliers from picked up high frequency signals on the lines for the supply voltages.
The signal $V_g$ is fed directly into the Y input of the left and the negated Y input of the right multiplier, whereas the signal $V_f$ is fed into the Y input of the upper and the lower multiplier.
The signal $V_\lambda$ which is added to compensate for parasitic capacitance entries on the diagonal of the Laplacian is fed into the Y input of the right multiplier.
The inputs that are not indicated in the schematic are grounded in both variants.}
\label{fig:Floquet-SSH-detailed}
\end{figure}

\begin{figure}[p]
\centering
\includegraphics[width=1.\linewidth]{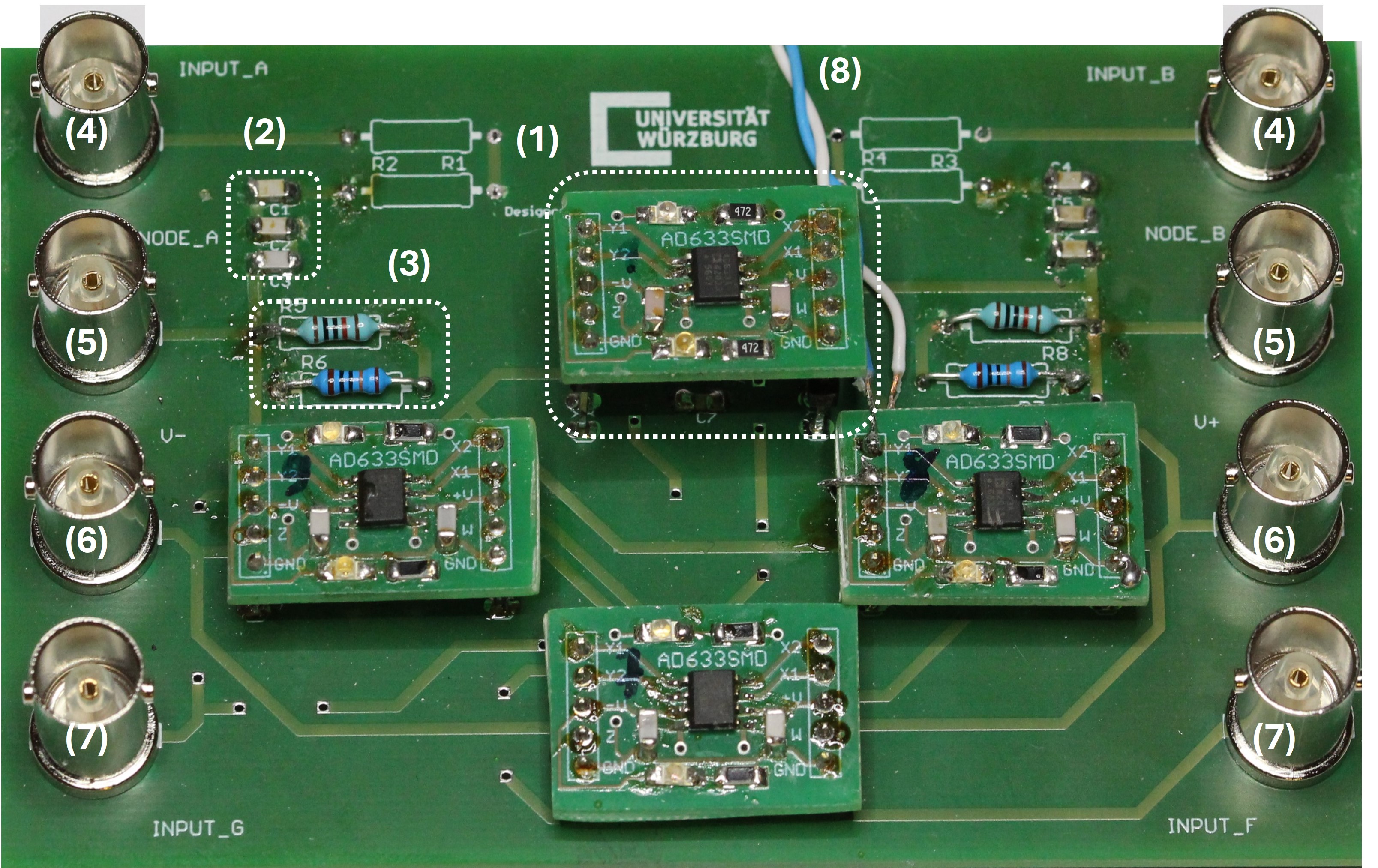}
\caption{
Circuit board of variant (b) of the Floquet SSH board.
\textbf{(1):}
Multiplier of type AD633 on top of an extra board.
On these extra boards light emitting diodes are placed to indicate working power supply for the multipliers.
Additionally, capacitors connect the supply lines to ground to decouple the supply lines from AC signals which they might have picked up.
Directly beneath the elevated boards, the capacitors connected to the outputs W of the multipliers are located. 
\textbf{(2):}
Three capacitors with a capacitance of $10~\mu\text{F}$ in series over which the input signals are fed into node $A$ and the integral of the current flowing into node $A$ is measured.
% The cables that lead to the FEMTO voltage amplifiers were soldered to the capacitors on the back of the board.
The cables that lead to the FEMTO voltage amplifiers were soldered to the back of the board.
\textbf{(3):}
Resistors in series with a total resistance of $110~\Omega$ to ground over which the voltage at node $A$ was measured using Tektronix TPP0250 probes attached to a Tektronix 4 Series mixed signal oscilloscope.
\textbf{(4):}
%Bayonet-Neill-Concelman 
BNC connector inputs for the signals into node $A$ (left) and node $B$ (right).
\textbf{(5):}
BNC connector inputs to node $A$ (left) and node $B$ (right) for measuring at node $A$ and node $B$ are obsolete since the voltages at the nodes are measured over the resistors to ground as described in (3).
\textbf{(6):}
BNC connector inputs for the supply voltages, negative (left) and positive (right).
\textbf{(7):}
BNC connector inputs for the signals $V_g$ (left) and $V_f$ (right).
\textbf{(8):}
Twisted pair cable to feed the signal $\lambda$ into the input $\text{Y}_1$ of the right multiplier. 
}
\label{fig:Floquet-SSH-board}
\end{figure}

Detailed circuit diagrams of the variants of implementation of the Floquet SSH chain are presented in Fig.~\ref{fig:Floquet-SSH-detailed}.
A picture of the physical board of variant (b) is presented in Fig.~\ref{fig:Floquet-SSH-board}.
Both variants of the Floquet SSH circuit consist of four analog multipliers of type AD633 by Analog Devices Inc, which are assembled on a breadboard (variant (a)) or custom printed circuit board (PCB) (variant (b)) together with all the other circuit components mentioned.
The outputs W of the upper multiplier and the left multiplier are connected to capacitors $C_1$ of type Murata GR442QR73D102KW01L with a nominal capacitance of 1\,nF.
The X inputs of the upper and the left multipliers are coupled to node $A$ as well as the capacitors $C_1$ of the lower and the left multiplier.
In contrast to variant (a), where the Z inputs are grounded, in variant (b) the Z inputs of the left and the lower multiplier are coupled to node $A$.
Additionally, grounded resistors are coupled to node $A$: In variant (a) two $100$~k$\Omega$ resistors of type Yageo MFR-25FTF52-100K in parallel and in variant (b) a $110$~k$\Omega$ resistor to ground consisting of a $100$~k$\Omega$ resistor of type Yageo MFR-25FTF52-100K and a $10$~k$\Omega$ resistor of type Yageo MFR-1WSFTE52-10K in series.
% These resistors in combination with the $10$~M$\Omega$ in parallel yield a total resistance of $99.1$~k$\Omega$ to ground for variant (b).

Also connected to node $A$ is the input signal $V_{\mathrm{in},A}$, which is either grounded or in the case of variant (a) a sine wave with an amplitude of $500~\text{mVpp}$ and a frequency of $k \cdot 10~\text{kHz}$ with $k\in\{-10, ..., 9\}$, whereas for variant (b) it is either grounded or a Gaussian pulse of the form 
\begin{align}
    250~\text{mV}\cdot\exp\left(-\frac{1}{16}\left(2\pi\left(t-5~\text{ms}\right)\cdot 10~\text{kHz}\right)^{2}\right)\cdot\sin\left(2\pi kt\cdot 10~\text{kHz} +  \pi/4\right) \quad \text{with} \quad k\in\{-10, ..., 9\}.
\end{align}
The signal enters the circuit in the case of variant (a) via a resistor of $500~\Omega$ of type Vishay RN60D5000FB14, over which the input currents are measured with oscilloscopes of type PicoScope 4824A and a $3.3$~nF capacitor $C_{\mathrm{in}}$, consisting of three 10\,nF capacitors of type Murata GCM31C5C2J103FX03L in series.
For variant (b), the integral of the current flowing into the circuit was measured directly by the voltage drop over the $3.3$~nF capacitor with a Tektronix 4 Series mixed signal oscilloscope.
The voltage drop was first amplified by a differential voltage amplifier Femto model DLPVA-101-F-D.
The voltage on node $A$ was for variant (a) measured with a PicoScope 4824A and for variant (b) measured using a 10x-probe of type Tektronix TPP0250 250 MHz, connected to the Tektronix 4 Series mixed signal oscilloscope.%, further reducing paracitic capacitances in the circuit.

Node $B$ is likewise connected to the X inputs of the right and the lower multiplier, as well as to the capacitors at the outputs W of the upper and the right multiplier, a $110$~k$\Omega$ resistor to ground, and a $3.3$~nF capacitor to $V_{\mathrm{in},B}$. The specifications for the resistors and capacitors were the same for nodes $A$ and $B$.
The signal $V_f$ to the Y input of the lower and upper multipliers and the signal $V_g$ to the Y and the negated Y input of the left and right multipliers, respectively, as well as the inputs at $V_{\mathrm{in},A}$ and $V_{\mathrm{in},B}$ were generated by function generators of the type Keysight 33500B.
% To guarantee the same time base for the function generators in use as well as the oscilloscopes a Stanford Research Systems Model SG384 signal generator was used.
The timebases of the function generators and the Tektronix oscilloscopes were synchronized with a common 10\,MHz reference clock provided by a Stanford Research Systems SG384 function generator.

By tuning the voltages $V_f$ and $V_g$, the coupling parameters $u,v$ can be varied. They are connected by the relation
\begin{align}
    \left(\begin{array}{cc}
v\sin \Omega t & u+v\cos \Omega t \\ 
u+v\cos \Omega t & -v\sin \Omega t
\end{array} \right)=\frac{1}{10\mathrm{V}}\left(\begin{array}{cc}
V_g & V_f \\ 
V_f & -V_g
\end{array} \right).
\end{align}
To compensate for parasitic capacitances in the experimental setup on the diagonal entries of this matrix, an additional offset $\lambda$ was introduced in variant (b) and added to the signal $V_g$ as an offset of $540$~mV %541mV
and further an offset of $1.08$~V to the Y input of the right multiplier causing an overall offset of $\lambda$ on both diagonal entries of the matrix above.
The supply voltages of $15~\text{V}$ were generated by a Keysight B2962B power source.
To shield the multipliers from high frequency signals that the supply voltage lines might pick up, the supply voltage inputs of the multipliers each had an additional $1~\mu \text{F}$ capacitor of type Murata GRM55DR72D105KW01L attached to ground.
\vspace*{2cm}	
       
\begin{thebibliography}{45}%
\makeatletter
\providecommand \@ifxundefined [1]{%
 \@ifx{#1\undefined}
}%
\providecommand \@ifnum [1]{%
 \ifnum #1\expandafter \@firstoftwo
 \else \expandafter \@secondoftwo
 \fi
}%
\providecommand \@ifx [1]{%
 \ifx #1\expandafter \@firstoftwo
 \else \expandafter \@secondoftwo
 \fi
}%
\providecommand \natexlab [1]{#1}%
\providecommand \enquote  [1]{``#1''}%
\providecommand \bibnamefont  [1]{#1}%
\providecommand \bibfnamefont [1]{#1}%
\providecommand \citenamefont [1]{#1}%
\providecommand \href@noop [0]{\@secondoftwo}%
\providecommand \href [0]{\begingroup \@sanitize@url \@href}%
\providecommand \@href[1]{\@@startlink{#1}\@@href}%
\providecommand \@@href[1]{\endgroup#1\@@endlink}%
\providecommand \@sanitize@url [0]{\catcode `\\12\catcode `\$12\catcode `\&12\catcode `\#12\catcode `\^12\catcode `\_12\catcode `\%12\relax}%
\providecommand \@@startlink[1]{}%
\providecommand \@@endlink[0]{}%
\providecommand \url  [0]{\begingroup\@sanitize@url \@url }%
\providecommand \@url [1]{\endgroup\@href {#1}{\urlprefix }}%
\providecommand \urlprefix  [0]{URL }%
\providecommand \Eprint [0]{\href }%
\providecommand \doibase [0]{http://dx.doi.org/}%
\providecommand \selectlanguage [0]{\@gobble}%
\providecommand \bibinfo  [0]{\@secondoftwo}%
\providecommand \bibfield  [0]{\@secondoftwo}%
\providecommand \translation [1]{[#1]}%
\providecommand \BibitemOpen [0]{}%
\providecommand \bibitemStop [0]{}%
\providecommand \bibitemNoStop [0]{.\EOS\space}%
\providecommand \EOS [0]{\spacefactor3000\relax}%
\providecommand \BibitemShut  [1]{\csname bibitem#1\endcsname}%
\let\auto@bib@innerbib\@empty
%</preamble>
\bibitem [{\citenamefont {Jackiw}\ and\ \citenamefont {Rebbi}(1976)}]{PhysRevD.13.3398}%
  \BibitemOpen
  \bibfield  {author} {\bibinfo {author} {\bibfnamefont {R.}~\bibnamefont {Jackiw}}\ and\ \bibinfo {author} {\bibfnamefont {C.}~\bibnamefont {Rebbi}},\ }\href {\doibase 10.1103/PhysRevD.13.3398} {\bibfield  {journal} {\bibinfo  {journal} {Phys. Rev. D}\ }\textbf {\bibinfo {volume} {13}},\ \bibinfo {pages} {3398} (\bibinfo {year} {1976})}\BibitemShut {NoStop}%
\bibitem [{\citenamefont {Su}\ \emph {et~al.}(1979)\citenamefont {Su}, \citenamefont {Schrieffer},\ and\ \citenamefont {Heeger}}]{Su1979solitons}%
  \BibitemOpen
  \bibfield  {author} {\bibinfo {author} {\bibfnamefont {W.~P.}\ \bibnamefont {Su}}, \bibinfo {author} {\bibfnamefont {J.~R.}\ \bibnamefont {Schrieffer}}, \ and\ \bibinfo {author} {\bibfnamefont {A.~J.}\ \bibnamefont {Heeger}},\ }\href {\doibase 10.1103/PhysRevLett.42.1698} {\bibfield  {journal} {\bibinfo  {journal} {Phys. Rev. Lett.}\ }\textbf {\bibinfo {volume} {42}},\ \bibinfo {pages} {1698} (\bibinfo {year} {1979})}\BibitemShut {NoStop}%
\bibitem [{\citenamefont {Halperin}(1982)}]{PhysRevB.25.2185}%
  \BibitemOpen
  \bibfield  {author} {\bibinfo {author} {\bibfnamefont {B.~I.}\ \bibnamefont {Halperin}},\ }\href {\doibase 10.1103/PhysRevB.25.2185} {\bibfield  {journal} {\bibinfo  {journal} {Phys. Rev. B}\ }\textbf {\bibinfo {volume} {25}},\ \bibinfo {pages} {2185} (\bibinfo {year} {1982})}\BibitemShut {NoStop}%
\bibitem [{\citenamefont {Volovik}(2009)}]{volovik2009universe}%
  \BibitemOpen
  \bibfield  {author} {\bibinfo {author} {\bibfnamefont {G.~E.}\ \bibnamefont {Volovik}},\ }\href {\doibase 10.1093/acprof:oso/9780199564842.001.0001} {\emph {\bibinfo {title} {{The Universe in a Helium Droplet}}}}\ (\bibinfo  {publisher} {Oxford University Press},\ \bibinfo {year} {2009})\BibitemShut {NoStop}%
\bibitem [{\citenamefont {Hasan}\ and\ \citenamefont {Kane}(2010)}]{Hasan2010}%
  \BibitemOpen
  \bibfield  {author} {\bibinfo {author} {\bibfnamefont {M.~Z.}\ \bibnamefont {Hasan}}\ and\ \bibinfo {author} {\bibfnamefont {C.~L.}\ \bibnamefont {Kane}},\ }\href {\doibase 10.1103/RevModPhys.82.3045} {\bibfield  {journal} {\bibinfo  {journal} {Rev. Mod. Phys.}\ }\textbf {\bibinfo {volume} {82}},\ \bibinfo {pages} {3045} (\bibinfo {year} {2010})}\BibitemShut {NoStop}%
\bibitem [{\citenamefont {Dutt}\ \emph {et~al.}(2022)\citenamefont {Dutt}, \citenamefont {Yuan}, \citenamefont {Yang}, \citenamefont {Wang}, \citenamefont {Buddhiraju}, \citenamefont {Vu{\v{c}}kovi{\'{c}}},\ and\ \citenamefont {Fan}}]{Dutt2022creating}%
  \BibitemOpen
  \bibfield  {author} {\bibinfo {author} {\bibfnamefont {A.}~\bibnamefont {Dutt}}, \bibinfo {author} {\bibfnamefont {L.}~\bibnamefont {Yuan}}, \bibinfo {author} {\bibfnamefont {K.~Y.}\ \bibnamefont {Yang}}, \bibinfo {author} {\bibfnamefont {K.}~\bibnamefont {Wang}}, \bibinfo {author} {\bibfnamefont {S.}~\bibnamefont {Buddhiraju}}, \bibinfo {author} {\bibfnamefont {J.}~\bibnamefont {Vu{\v{c}}kovi{\'{c}}}}, \ and\ \bibinfo {author} {\bibfnamefont {S.}~\bibnamefont {Fan}},\ }\href {\doibase 10.1038/s41467-022-31140-7} {\bibfield  {journal} {\bibinfo  {journal} {Nat. Commun.}\ }\textbf {\bibinfo {volume} {13}},\ \bibinfo {pages} {3377} (\bibinfo {year} {2022})}\BibitemShut {NoStop}%
\bibitem [{\citenamefont {Ozawa}\ and\ \citenamefont {Price}(2019)}]{Ozawa2019}%
  \BibitemOpen
  \bibfield  {author} {\bibinfo {author} {\bibfnamefont {T.}~\bibnamefont {Ozawa}}\ and\ \bibinfo {author} {\bibfnamefont {H.~M.}\ \bibnamefont {Price}},\ }\href {\doibase 10.1038/s42254-019-0045-3} {\bibfield  {journal} {\bibinfo  {journal} {Nat. Rev. Phys.}\ }\textbf {\bibinfo {volume} {1}},\ \bibinfo {pages} {349} (\bibinfo {year} {2019})}\BibitemShut {NoStop}%
\bibitem [{\citenamefont {Yuan}\ \emph {et~al.}(2018)\citenamefont {Yuan}, \citenamefont {Lin}, \citenamefont {Xiao},\ and\ \citenamefont {Fan}}]{Yuan18}%
  \BibitemOpen
  \bibfield  {author} {\bibinfo {author} {\bibfnamefont {L.}~\bibnamefont {Yuan}}, \bibinfo {author} {\bibfnamefont {Q.}~\bibnamefont {Lin}}, \bibinfo {author} {\bibfnamefont {M.}~\bibnamefont {Xiao}}, \ and\ \bibinfo {author} {\bibfnamefont {S.}~\bibnamefont {Fan}},\ }\href {\doibase 10.1364/OPTICA.5.001396} {\bibfield  {journal} {\bibinfo  {journal} {Optica}\ }\textbf {\bibinfo {volume} {5}},\ \bibinfo {pages} {1396} (\bibinfo {year} {2018})}\BibitemShut {NoStop}%
\bibitem [{\citenamefont {Lustig}\ and\ \citenamefont {Segev}(2021)}]{Lustig21TopoPhot}%
  \BibitemOpen
  \bibfield  {author} {\bibinfo {author} {\bibfnamefont {E.}~\bibnamefont {Lustig}}\ and\ \bibinfo {author} {\bibfnamefont {M.}~\bibnamefont {Segev}},\ }\href {\doibase 10.1364/AOP.418074} {\bibfield  {journal} {\bibinfo  {journal} {Adv. Opt. Photon.}\ }\textbf {\bibinfo {volume} {13}},\ \bibinfo {pages} {426} (\bibinfo {year} {2021})}\BibitemShut {NoStop}%
\bibitem [{\citenamefont {Yang}\ \emph {et~al.}(2022)\citenamefont {Yang}, \citenamefont {Xu}, \citenamefont {Li},\ and\ \citenamefont {Guo}}]{Yang2022cavityrev}%
  \BibitemOpen
  \bibfield  {author} {\bibinfo {author} {\bibfnamefont {M.}~\bibnamefont {Yang}}, \bibinfo {author} {\bibfnamefont {J.-S.}\ \bibnamefont {Xu}}, \bibinfo {author} {\bibfnamefont {C.-F.}\ \bibnamefont {Li}}, \ and\ \bibinfo {author} {\bibfnamefont {G.-C.}\ \bibnamefont {Guo}},\ }\href {\doibase 10.1007/s44214-022-00015-9} {\bibfield  {journal} {\bibinfo  {journal} {Quantum Front.}\ }\textbf {\bibinfo {volume} {1}},\ \bibinfo {pages} {10} (\bibinfo {year} {2022})}\BibitemShut {NoStop}%
\bibitem [{\citenamefont {Ehrhardt}\ \emph {et~al.}(2023)\citenamefont {Ehrhardt}, \citenamefont {Weidemann}, \citenamefont {Maczewsky}, \citenamefont {Heinrich},\ and\ \citenamefont {Szameit}}]{Ehrhardt2023}%
  \BibitemOpen
  \bibfield  {author} {\bibinfo {author} {\bibfnamefont {M.}~\bibnamefont {Ehrhardt}}, \bibinfo {author} {\bibfnamefont {S.}~\bibnamefont {Weidemann}}, \bibinfo {author} {\bibfnamefont {L.~J.}\ \bibnamefont {Maczewsky}}, \bibinfo {author} {\bibfnamefont {M.}~\bibnamefont {Heinrich}}, \ and\ \bibinfo {author} {\bibfnamefont {A.}~\bibnamefont {Szameit}},\ }\href {\doibase https://doi.org/10.1002/lpor.202200518} {\bibfield  {journal} {\bibinfo  {journal} {Laser Photonics Rev.}\ }\textbf {\bibinfo {volume} {17}},\ \bibinfo {pages} {2200518} (\bibinfo {year} {2023})}\BibitemShut {NoStop}%
\bibitem [{\citenamefont {Lustig}\ \emph {et~al.}(2019)\citenamefont {Lustig}, \citenamefont {Weimann}, \citenamefont {Plotnik}, \citenamefont {Lumer}, \citenamefont {Bandres}, \citenamefont {Szameit},\ and\ \citenamefont {Segev}}]{Lustig2019}%
  \BibitemOpen
  \bibfield  {author} {\bibinfo {author} {\bibfnamefont {E.}~\bibnamefont {Lustig}}, \bibinfo {author} {\bibfnamefont {S.}~\bibnamefont {Weimann}}, \bibinfo {author} {\bibfnamefont {Y.}~\bibnamefont {Plotnik}}, \bibinfo {author} {\bibfnamefont {Y.}~\bibnamefont {Lumer}}, \bibinfo {author} {\bibfnamefont {M.~A.}\ \bibnamefont {Bandres}}, \bibinfo {author} {\bibfnamefont {A.}~\bibnamefont {Szameit}}, \ and\ \bibinfo {author} {\bibfnamefont {M.}~\bibnamefont {Segev}},\ }\href {\doibase 10.1038/s41586-019-0943-7} {\bibfield  {journal} {\bibinfo  {journal} {Nature}\ }\textbf {\bibinfo {volume} {567}},\ \bibinfo {pages} {356} (\bibinfo {year} {2019})}\BibitemShut {NoStop}%
\bibitem [{\citenamefont {Pirruccio}\ and\ \citenamefont {Naumis}(2022)}]{Pirrucio2022cavities}%
  \BibitemOpen
  \bibfield  {author} {\bibinfo {author} {\bibfnamefont {G.}~\bibnamefont {Pirruccio}}\ and\ \bibinfo {author} {\bibfnamefont {G.~G.}\ \bibnamefont {Naumis}},\ }\href {\doibase 10.1103/PhysRevB.106.035155} {\bibfield  {journal} {\bibinfo  {journal} {Phys. Rev. B}\ }\textbf {\bibinfo {volume} {106}},\ \bibinfo {pages} {035155} (\bibinfo {year} {2022})}\BibitemShut {NoStop}%
\bibitem [{\citenamefont {Dutt}\ \emph {et~al.}(2020)\citenamefont {Dutt}, \citenamefont {Lin}, \citenamefont {Yuan}, \citenamefont {Minkov}, \citenamefont {Xiao},\ and\ \citenamefont {Fan}}]{Dutt2020}%
  \BibitemOpen
  \bibfield  {author} {\bibinfo {author} {\bibfnamefont {A.}~\bibnamefont {Dutt}}, \bibinfo {author} {\bibfnamefont {Q.}~\bibnamefont {Lin}}, \bibinfo {author} {\bibfnamefont {L.}~\bibnamefont {Yuan}}, \bibinfo {author} {\bibfnamefont {M.}~\bibnamefont {Minkov}}, \bibinfo {author} {\bibfnamefont {M.}~\bibnamefont {Xiao}}, \ and\ \bibinfo {author} {\bibfnamefont {S.}~\bibnamefont {Fan}},\ }\href {\doibase 10.1126/science.aaz3071} {\bibfield  {journal} {\bibinfo  {journal} {Science}\ }\textbf {\bibinfo {volume} {367}},\ \bibinfo {pages} {59} (\bibinfo {year} {2020})}\BibitemShut {NoStop}%
\bibitem [{\citenamefont {Yuan}\ \emph {et~al.}(2021)\citenamefont {Yuan}, \citenamefont {Dutt},\ and\ \citenamefont {Fan}}]{Yuan2021}%
  \BibitemOpen
  \bibfield  {author} {\bibinfo {author} {\bibfnamefont {L.}~\bibnamefont {Yuan}}, \bibinfo {author} {\bibfnamefont {A.}~\bibnamefont {Dutt}}, \ and\ \bibinfo {author} {\bibfnamefont {S.}~\bibnamefont {Fan}},\ }\href {\doibase 10.1063/5.0056359} {\bibfield  {journal} {\bibinfo  {journal} {APL Photonics}\ }\textbf {\bibinfo {volume} {6}},\ \bibinfo {pages} {071102} (\bibinfo {year} {2021})}\BibitemShut {NoStop}%
\bibitem [{\citenamefont {Li}\ \emph {et~al.}(2021)\citenamefont {Li}, \citenamefont {Zheng}, \citenamefont {Dutt}, \citenamefont {Yu}, \citenamefont {Shan}, \citenamefont {Liu}, \citenamefont {Yuan}, \citenamefont {Fan},\ and\ \citenamefont {Chen}}]{Li2021}%
  \BibitemOpen
  \bibfield  {author} {\bibinfo {author} {\bibfnamefont {G.}~\bibnamefont {Li}}, \bibinfo {author} {\bibfnamefont {Y.}~\bibnamefont {Zheng}}, \bibinfo {author} {\bibfnamefont {A.}~\bibnamefont {Dutt}}, \bibinfo {author} {\bibfnamefont {D.}~\bibnamefont {Yu}}, \bibinfo {author} {\bibfnamefont {Q.}~\bibnamefont {Shan}}, \bibinfo {author} {\bibfnamefont {S.}~\bibnamefont {Liu}}, \bibinfo {author} {\bibfnamefont {L.}~\bibnamefont {Yuan}}, \bibinfo {author} {\bibfnamefont {S.}~\bibnamefont {Fan}}, \ and\ \bibinfo {author} {\bibfnamefont {X.}~\bibnamefont {Chen}},\ }\href {\doibase 10.1126/sciadv.abe4335} {\bibfield  {journal} {\bibinfo  {journal} {Sci. Adv.}\ }\textbf {\bibinfo {volume} {7}},\ \bibinfo {pages} {eabe4335} (\bibinfo {year} {2021})}\BibitemShut {NoStop}%
\bibitem [{\citenamefont {Cheng}\ \emph {et~al.}(2023)\citenamefont {Cheng}, \citenamefont {Lustig}, \citenamefont {Wang},\ and\ \citenamefont {Fan}}]{Cheng2023syndimband}%
  \BibitemOpen
  \bibfield  {author} {\bibinfo {author} {\bibfnamefont {D.}~\bibnamefont {Cheng}}, \bibinfo {author} {\bibfnamefont {E.}~\bibnamefont {Lustig}}, \bibinfo {author} {\bibfnamefont {K.}~\bibnamefont {Wang}}, \ and\ \bibinfo {author} {\bibfnamefont {S.}~\bibnamefont {Fan}},\ }\href {\doibase 10.1038/s41377-023-01196-1} {\bibfield  {journal} {\bibinfo  {journal} {Light Sci. Appl.}\ }\textbf {\bibinfo {volume} {12}},\ \bibinfo {pages} {158} (\bibinfo {year} {2023})}\BibitemShut {NoStop}%
\bibitem [{\citenamefont {Long}\ \emph {et~al.}(2023)\citenamefont {Long}, \citenamefont {Wang}, \citenamefont {Dutt},\ and\ \citenamefont {Fan}}]{Long23timebound}%
  \BibitemOpen
  \bibfield  {author} {\bibinfo {author} {\bibfnamefont {O.~Y.}\ \bibnamefont {Long}}, \bibinfo {author} {\bibfnamefont {K.}~\bibnamefont {Wang}}, \bibinfo {author} {\bibfnamefont {A.}~\bibnamefont {Dutt}}, \ and\ \bibinfo {author} {\bibfnamefont {S.}~\bibnamefont {Fan}},\ }\href {\doibase 10.1103/PhysRevResearch.5.L012046} {\bibfield  {journal} {\bibinfo  {journal} {Phys. Rev. Res.}\ }\textbf {\bibinfo {volume} {5}},\ \bibinfo {pages} {L012046} (\bibinfo {year} {2023})}\BibitemShut {NoStop}%
\bibitem [{\citenamefont {Liu}\ \emph {et~al.}(2023)\citenamefont {Liu}, \citenamefont {Zheng}, \citenamefont {Qin}, \citenamefont {Wang},\ and\ \citenamefont {Lu}}]{Liu2023blochsyndimwaveg}%
  \BibitemOpen
  \bibfield  {author} {\bibinfo {author} {\bibfnamefont {Z.}~\bibnamefont {Liu}}, \bibinfo {author} {\bibfnamefont {L.}~\bibnamefont {Zheng}}, \bibinfo {author} {\bibfnamefont {C.}~\bibnamefont {Qin}}, \bibinfo {author} {\bibfnamefont {B.}~\bibnamefont {Wang}}, \ and\ \bibinfo {author} {\bibfnamefont {P.}~\bibnamefont {Lu}},\ }\href {\doibase 10.1364/OL.491680} {\bibfield  {journal} {\bibinfo  {journal} {Opt. Lett.}\ }\textbf {\bibinfo {volume} {48}},\ \bibinfo {pages} {3163} (\bibinfo {year} {2023})}\BibitemShut {NoStop}%
\bibitem [{\citenamefont {Nemirovsky}\ \emph {et~al.}(2021)\citenamefont {Nemirovsky}, \citenamefont {Cohen}, \citenamefont {Lumer}, \citenamefont {Lustig},\ and\ \citenamefont {Segev}}]{Nemirovsky2021}%
  \BibitemOpen
  \bibfield  {author} {\bibinfo {author} {\bibfnamefont {L.}~\bibnamefont {Nemirovsky}}, \bibinfo {author} {\bibfnamefont {M.-I.}\ \bibnamefont {Cohen}}, \bibinfo {author} {\bibfnamefont {Y.}~\bibnamefont {Lumer}}, \bibinfo {author} {\bibfnamefont {E.}~\bibnamefont {Lustig}}, \ and\ \bibinfo {author} {\bibfnamefont {M.}~\bibnamefont {Segev}},\ }\href {\doibase 10.1103/PhysRevLett.127.093901} {\bibfield  {journal} {\bibinfo  {journal} {Phys. Rev. Lett.}\ }\textbf {\bibinfo {volume} {127}},\ \bibinfo {pages} {093901} (\bibinfo {year} {2021})}\BibitemShut {NoStop}%
\bibitem [{\citenamefont {Grinberg}\ \emph {et~al.}(2020)\citenamefont {Grinberg}, \citenamefont {Lin}, \citenamefont {Harris}, \citenamefont {Benalcazar}, \citenamefont {Peterson}, \citenamefont {Hughes},\ and\ \citenamefont {Bahl}}]{Grinberg2020}%
  \BibitemOpen
  \bibfield  {author} {\bibinfo {author} {\bibfnamefont {I.~H.}\ \bibnamefont {Grinberg}}, \bibinfo {author} {\bibfnamefont {M.}~\bibnamefont {Lin}}, \bibinfo {author} {\bibfnamefont {C.}~\bibnamefont {Harris}}, \bibinfo {author} {\bibfnamefont {W.~A.}\ \bibnamefont {Benalcazar}}, \bibinfo {author} {\bibfnamefont {C.~W.}\ \bibnamefont {Peterson}}, \bibinfo {author} {\bibfnamefont {T.~L.}\ \bibnamefont {Hughes}}, \ and\ \bibinfo {author} {\bibfnamefont {G.}~\bibnamefont {Bahl}},\ }\href {\doibase 10.1038/s41467-020-14804-0} {\bibfield  {journal} {\bibinfo  {journal} {Nat. Commun.}\ }\textbf {\bibinfo {volume} {11}},\ \bibinfo {pages} {974} (\bibinfo {year} {2020})}\BibitemShut {NoStop}%
\bibitem [{\citenamefont {Oliver}\ \emph {et~al.}(2023)\citenamefont {Oliver}, \citenamefont {Smith}, \citenamefont {Easton}, \citenamefont {Salerno}, \citenamefont {Guarrera}, \citenamefont {Goldman}, \citenamefont {Barontini},\ and\ \citenamefont {Price}}]{Oliver2023}%
  \BibitemOpen
  \bibfield  {author} {\bibinfo {author} {\bibfnamefont {C.}~\bibnamefont {Oliver}}, \bibinfo {author} {\bibfnamefont {A.}~\bibnamefont {Smith}}, \bibinfo {author} {\bibfnamefont {T.}~\bibnamefont {Easton}}, \bibinfo {author} {\bibfnamefont {G.}~\bibnamefont {Salerno}}, \bibinfo {author} {\bibfnamefont {V.}~\bibnamefont {Guarrera}}, \bibinfo {author} {\bibfnamefont {N.}~\bibnamefont {Goldman}}, \bibinfo {author} {\bibfnamefont {G.}~\bibnamefont {Barontini}}, \ and\ \bibinfo {author} {\bibfnamefont {H.~M.}\ \bibnamefont {Price}},\ }\href {\doibase 10.1103/PhysRevResearch.5.033001} {\bibfield  {journal} {\bibinfo  {journal} {Phys. Rev. Res.}\ }\textbf {\bibinfo {volume} {5}},\ \bibinfo {pages} {033001} (\bibinfo {year} {2023})}\BibitemShut {NoStop}%
\bibitem [{\citenamefont {Rechtsman}\ \emph {et~al.}(2013)\citenamefont {Rechtsman}, \citenamefont {Zeuner}, \citenamefont {Plotnik}, \citenamefont {Lumer}, \citenamefont {Podolsky}, \citenamefont {Dreisow}, \citenamefont {Nolte}, \citenamefont {Segev},\ and\ \citenamefont {Szameit}}]{Rechtsman2013PFTI}%
  \BibitemOpen
  \bibfield  {author} {\bibinfo {author} {\bibfnamefont {M.~C.}\ \bibnamefont {Rechtsman}}, \bibinfo {author} {\bibfnamefont {J.~M.}\ \bibnamefont {Zeuner}}, \bibinfo {author} {\bibfnamefont {Y.}~\bibnamefont {Plotnik}}, \bibinfo {author} {\bibfnamefont {Y.}~\bibnamefont {Lumer}}, \bibinfo {author} {\bibfnamefont {D.}~\bibnamefont {Podolsky}}, \bibinfo {author} {\bibfnamefont {F.}~\bibnamefont {Dreisow}}, \bibinfo {author} {\bibfnamefont {S.}~\bibnamefont {Nolte}}, \bibinfo {author} {\bibfnamefont {M.}~\bibnamefont {Segev}}, \ and\ \bibinfo {author} {\bibfnamefont {A.}~\bibnamefont {Szameit}},\ }\href {\doibase 10.1038/nature12066} {\bibfield  {journal} {\bibinfo  {journal} {Nature}\ }\textbf {\bibinfo {volume} {496}},\ \bibinfo {pages} {196} (\bibinfo {year} {2013})}\BibitemShut {NoStop}%
\bibitem [{\citenamefont {Rudner}\ \emph {et~al.}(2013)\citenamefont {Rudner}, \citenamefont {Lindner}, \citenamefont {Berg},\ and\ \citenamefont {Levin}}]{Rudner2013anomal}%
  \BibitemOpen
  \bibfield  {author} {\bibinfo {author} {\bibfnamefont {M.~S.}\ \bibnamefont {Rudner}}, \bibinfo {author} {\bibfnamefont {N.~H.}\ \bibnamefont {Lindner}}, \bibinfo {author} {\bibfnamefont {E.}~\bibnamefont {Berg}}, \ and\ \bibinfo {author} {\bibfnamefont {M.}~\bibnamefont {Levin}},\ }\href {\doibase 10.1103/PhysRevX.3.031005} {\bibfield  {journal} {\bibinfo  {journal} {Phys. Rev. X}\ }\textbf {\bibinfo {volume} {3}},\ \bibinfo {pages} {031005} (\bibinfo {year} {2013})}\BibitemShut {NoStop}%
\bibitem [{\citenamefont {Martin}\ \emph {et~al.}(2017)\citenamefont {Martin}, \citenamefont {Refael},\ and\ \citenamefont {Halperin}}]{Martin2017}%
  \BibitemOpen
  \bibfield  {author} {\bibinfo {author} {\bibfnamefont {I.}~\bibnamefont {Martin}}, \bibinfo {author} {\bibfnamefont {G.}~\bibnamefont {Refael}}, \ and\ \bibinfo {author} {\bibfnamefont {B.}~\bibnamefont {Halperin}},\ }\href {\doibase 10.1103/PhysRevX.7.041008} {\bibfield  {journal} {\bibinfo  {journal} {Phys. Rev. X}\ }\textbf {\bibinfo {volume} {7}},\ \bibinfo {pages} {041008} (\bibinfo {year} {2017})}\BibitemShut {NoStop}%
\bibitem [{\citenamefont {Nathan}\ \emph {et~al.}(2022)\citenamefont {Nathan}, \citenamefont {Martin},\ and\ \citenamefont {Refael}}]{Nathan2022}%
  \BibitemOpen
  \bibfield  {author} {\bibinfo {author} {\bibfnamefont {F.}~\bibnamefont {Nathan}}, \bibinfo {author} {\bibfnamefont {I.}~\bibnamefont {Martin}}, \ and\ \bibinfo {author} {\bibfnamefont {G.}~\bibnamefont {Refael}},\ }\href {\doibase 10.1103/PhysRevResearch.4.043060} {\bibfield  {journal} {\bibinfo  {journal} {Phys. Rev. Res.}\ }\textbf {\bibinfo {volume} {4}},\ \bibinfo {pages} {043060} (\bibinfo {year} {2022})}\BibitemShut {NoStop}%
\bibitem [{\citenamefont {Sridhar}\ \emph {et~al.}(2024)\citenamefont {Sridhar}, \citenamefont {Ghosh}, \citenamefont {Srinivasan}, \citenamefont {Miller},\ and\ \citenamefont {Dutt}}]{Sridhar2024}%
  \BibitemOpen
  \bibfield  {author} {\bibinfo {author} {\bibfnamefont {S.~K.}\ \bibnamefont {Sridhar}}, \bibinfo {author} {\bibfnamefont {S.}~\bibnamefont {Ghosh}}, \bibinfo {author} {\bibfnamefont {D.}~\bibnamefont {Srinivasan}}, \bibinfo {author} {\bibfnamefont {A.~R.}\ \bibnamefont {Miller}}, \ and\ \bibinfo {author} {\bibfnamefont {A.}~\bibnamefont {Dutt}},\ }\href {\doibase 10.1038/s41567-024-02413-3} {\bibfield  {journal} {\bibinfo  {journal} {Nature Physics}\ }\textbf {\bibinfo {volume} {20}},\ \bibinfo {pages} {843} (\bibinfo {year} {2024})}\BibitemShut {NoStop}%
\bibitem [{\citenamefont {Baum}\ and\ \citenamefont {Refael}(2018)}]{Baum18memory}%
  \BibitemOpen
  \bibfield  {author} {\bibinfo {author} {\bibfnamefont {Y.}~\bibnamefont {Baum}}\ and\ \bibinfo {author} {\bibfnamefont {G.}~\bibnamefont {Refael}},\ }\href {\doibase 10.1103/PhysRevLett.120.106402} {\bibfield  {journal} {\bibinfo  {journal} {Phys. Rev. Lett.}\ }\textbf {\bibinfo {volume} {120}},\ \bibinfo {pages} {106402} (\bibinfo {year} {2018})}\BibitemShut {NoStop}%
\bibitem [{\citenamefont {Lee}\ \emph {et~al.}(2018)\citenamefont {Lee}, \citenamefont {Imhof}, \citenamefont {Berger}, \citenamefont {Bayer}, \citenamefont {Brehm}, \citenamefont {Molenkamp}, \citenamefont {Kiessling},\ and\ \citenamefont {Thomale}}]{Lee2018circuits}%
  \BibitemOpen
  \bibfield  {author} {\bibinfo {author} {\bibfnamefont {C.~H.}\ \bibnamefont {Lee}}, \bibinfo {author} {\bibfnamefont {S.}~\bibnamefont {Imhof}}, \bibinfo {author} {\bibfnamefont {C.}~\bibnamefont {Berger}}, \bibinfo {author} {\bibfnamefont {F.}~\bibnamefont {Bayer}}, \bibinfo {author} {\bibfnamefont {J.}~\bibnamefont {Brehm}}, \bibinfo {author} {\bibfnamefont {L.~W.}\ \bibnamefont {Molenkamp}}, \bibinfo {author} {\bibfnamefont {T.}~\bibnamefont {Kiessling}}, \ and\ \bibinfo {author} {\bibfnamefont {R.}~\bibnamefont {Thomale}},\ }\href {\doibase 10.1038/s42005-018-0035-2} {\bibfield  {journal} {\bibinfo  {journal} {Comm. Phys.}\ }\textbf {\bibinfo {volume} {1}},\ \bibinfo {pages} {39} (\bibinfo {year} {2018})}\BibitemShut {NoStop}%
\bibitem [{\citenamefont {Helbig}\ \emph {et~al.}(2019)\citenamefont {Helbig}, \citenamefont {Hofmann}, \citenamefont {Lee}, \citenamefont {Thomale}, \citenamefont {Imhof}, \citenamefont {Molenkamp},\ and\ \citenamefont {Kiessling}}]{Helbig2019}%
  \BibitemOpen
  \bibfield  {author} {\bibinfo {author} {\bibfnamefont {T.}~\bibnamefont {Helbig}}, \bibinfo {author} {\bibfnamefont {T.}~\bibnamefont {Hofmann}}, \bibinfo {author} {\bibfnamefont {C.~H.}\ \bibnamefont {Lee}}, \bibinfo {author} {\bibfnamefont {R.}~\bibnamefont {Thomale}}, \bibinfo {author} {\bibfnamefont {S.}~\bibnamefont {Imhof}}, \bibinfo {author} {\bibfnamefont {L.~W.}\ \bibnamefont {Molenkamp}}, \ and\ \bibinfo {author} {\bibfnamefont {T.}~\bibnamefont {Kiessling}},\ }\href {\doibase 10.1103/PhysRevB.99.161114} {\bibfield  {journal} {\bibinfo  {journal} {Phys. Rev. B}\ }\textbf {\bibinfo {volume} {99}},\ \bibinfo {pages} {161114} (\bibinfo {year} {2019})}\BibitemShut {NoStop}%
\bibitem [{\citenamefont {Ezawa}(2019)}]{Ezawa2019}%
  \BibitemOpen
  \bibfield  {author} {\bibinfo {author} {\bibfnamefont {M.}~\bibnamefont {Ezawa}},\ }\href {\doibase 10.1103/PhysRevB.100.081401} {\bibfield  {journal} {\bibinfo  {journal} {Phys. Rev. B}\ }\textbf {\bibinfo {volume} {100}},\ \bibinfo {pages} {081401} (\bibinfo {year} {2019})}\BibitemShut {NoStop}%
\bibitem [{\citenamefont {Olekhno}\ \emph {et~al.}(2020)\citenamefont {Olekhno}, \citenamefont {Kretov}, \citenamefont {Stepanenko}, \citenamefont {Ivanova}, \citenamefont {Yaroshenko}, \citenamefont {Puhtina}, \citenamefont {Filonov}, \citenamefont {Cappello}, \citenamefont {Matekovits},\ and\ \citenamefont {Gorlach}}]{Olekhno2020}%
  \BibitemOpen
  \bibfield  {author} {\bibinfo {author} {\bibfnamefont {N.~A.}\ \bibnamefont {Olekhno}}, \bibinfo {author} {\bibfnamefont {E.~I.}\ \bibnamefont {Kretov}}, \bibinfo {author} {\bibfnamefont {A.~A.}\ \bibnamefont {Stepanenko}}, \bibinfo {author} {\bibfnamefont {P.~A.}\ \bibnamefont {Ivanova}}, \bibinfo {author} {\bibfnamefont {V.~V.}\ \bibnamefont {Yaroshenko}}, \bibinfo {author} {\bibfnamefont {E.~M.}\ \bibnamefont {Puhtina}}, \bibinfo {author} {\bibfnamefont {D.~S.}\ \bibnamefont {Filonov}}, \bibinfo {author} {\bibfnamefont {B.}~\bibnamefont {Cappello}}, \bibinfo {author} {\bibfnamefont {L.}~\bibnamefont {Matekovits}}, \ and\ \bibinfo {author} {\bibfnamefont {M.~A.}\ \bibnamefont {Gorlach}},\ }\href {\doibase 10.1038/s41467-020-14994-7} {\bibfield  {journal} {\bibinfo  {journal} {Nat. Commun.}\ }\textbf {\bibinfo {volume} {11}},\ \bibinfo {pages} {1436} (\bibinfo {year} {2020})}\BibitemShut {NoStop}%
\bibitem [{\citenamefont {Wang}\ \emph {et~al.}(2020)\citenamefont {Wang}, \citenamefont {Price}, \citenamefont {Zhang},\ and\ \citenamefont {Chong}}]{Wang2020_1}%
  \BibitemOpen
  \bibfield  {author} {\bibinfo {author} {\bibfnamefont {Y.}~\bibnamefont {Wang}}, \bibinfo {author} {\bibfnamefont {H.~M.}\ \bibnamefont {Price}}, \bibinfo {author} {\bibfnamefont {B.}~\bibnamefont {Zhang}}, \ and\ \bibinfo {author} {\bibfnamefont {Y.~D.}\ \bibnamefont {Chong}},\ }\href {\doibase 10.1038/s41467-020-15940-3} {\bibfield  {journal} {\bibinfo  {journal} {Nat. Commun.}\ }\textbf {\bibinfo {volume} {11}},\ \bibinfo {pages} {2356} (\bibinfo {year} {2020})}\BibitemShut {NoStop}%
\bibitem [{\citenamefont {Helbig}\ \emph {et~al.}(2020)\citenamefont {Helbig}, \citenamefont {Hofmann}, \citenamefont {Imhof}, \citenamefont {Abdelghany}, \citenamefont {Kiessling}, \citenamefont {Molenkamp}, \citenamefont {Lee}, \citenamefont {Szameit}, \citenamefont {Greiter},\ and\ \citenamefont {Thomale}}]{Helbig2020}%
  \BibitemOpen
  \bibfield  {author} {\bibinfo {author} {\bibfnamefont {T.}~\bibnamefont {Helbig}}, \bibinfo {author} {\bibfnamefont {T.}~\bibnamefont {Hofmann}}, \bibinfo {author} {\bibfnamefont {S.}~\bibnamefont {Imhof}}, \bibinfo {author} {\bibfnamefont {M.}~\bibnamefont {Abdelghany}}, \bibinfo {author} {\bibfnamefont {T.}~\bibnamefont {Kiessling}}, \bibinfo {author} {\bibfnamefont {L.~W.}\ \bibnamefont {Molenkamp}}, \bibinfo {author} {\bibfnamefont {C.~H.}\ \bibnamefont {Lee}}, \bibinfo {author} {\bibfnamefont {A.}~\bibnamefont {Szameit}}, \bibinfo {author} {\bibfnamefont {M.}~\bibnamefont {Greiter}}, \ and\ \bibinfo {author} {\bibfnamefont {R.}~\bibnamefont {Thomale}},\ }\href {\doibase 10.1038/s41567-020-0922-9} {\bibfield  {journal} {\bibinfo  {journal} {Nat. Phys.}\ }\textbf {\bibinfo {volume} {16}},\ \bibinfo {pages} {747} (\bibinfo {year} {2020})}\BibitemShut {NoStop}%
\bibitem [{\citenamefont {Hofmann}\ \emph {et~al.}(2020)\citenamefont {Hofmann}, \citenamefont {Helbig}, \citenamefont {Schindler}, \citenamefont {Salgo}, \citenamefont {Brzezi\ifmmode~\acute{n}\else \'{n}\fi{}ska}, \citenamefont {Greiter}, \citenamefont {Kiessling}, \citenamefont {Wolf}, \citenamefont {Vollhardt}, \citenamefont {Kaba\ifmmode~\check{s}\else \v{s}\fi{}i}, \citenamefont {Lee}, \citenamefont {Bilu\ifmmode \check{s}\else \v{s}\fi{}i\ifmmode~\acute{c}\else \'{c}\fi{}}, \citenamefont {Thomale},\ and\ \citenamefont {Neupert}}]{Hofmann2020}%
  \BibitemOpen
  \bibfield  {author} {\bibinfo {author} {\bibfnamefont {T.}~\bibnamefont {Hofmann}}, \bibinfo {author} {\bibfnamefont {T.}~\bibnamefont {Helbig}}, \bibinfo {author} {\bibfnamefont {F.}~\bibnamefont {Schindler}}, \bibinfo {author} {\bibfnamefont {N.}~\bibnamefont {Salgo}}, \bibinfo {author} {\bibfnamefont {M.}~\bibnamefont {Brzezi\ifmmode~\acute{n}\else \'{n}\fi{}ska}}, \bibinfo {author} {\bibfnamefont {M.}~\bibnamefont {Greiter}}, \bibinfo {author} {\bibfnamefont {T.}~\bibnamefont {Kiessling}}, \bibinfo {author} {\bibfnamefont {D.}~\bibnamefont {Wolf}}, \bibinfo {author} {\bibfnamefont {A.}~\bibnamefont {Vollhardt}}, \bibinfo {author} {\bibfnamefont {A.}~\bibnamefont {Kaba\ifmmode~\check{s}\else \v{s}\fi{}i}}, \bibinfo {author} {\bibfnamefont {C.~H.}\ \bibnamefont {Lee}}, \bibinfo {author} {\bibfnamefont {A.}~\bibnamefont {Bilu\ifmmode \check{s}\else \v{s}\fi{}i\ifmmode~\acute{c}\else \'{c}\fi{}}}, \bibinfo {author} {\bibfnamefont {R.}~\bibnamefont {Thomale}}, \ and\ \bibinfo {author} {\bibfnamefont
  {T.}~\bibnamefont {Neupert}},\ }\href {\doibase 10.1103/PhysRevResearch.2.023265} {\bibfield  {journal} {\bibinfo  {journal} {Phys. Rev. Res.}\ }\textbf {\bibinfo {volume} {2}},\ \bibinfo {pages} {023265} (\bibinfo {year} {2020})}\BibitemShut {NoStop}%
\bibitem [{\citenamefont {Stegmaier}\ \emph {et~al.}(2021)\citenamefont {Stegmaier}, \citenamefont {Imhof}, \citenamefont {Helbig}, \citenamefont {Hofmann}, \citenamefont {Lee}, \citenamefont {Kremer}, \citenamefont {Fritzsche}, \citenamefont {Feichtner}, \citenamefont {Klembt}, \citenamefont {H\"ofling}, \citenamefont {Boettcher}, \citenamefont {Fulga}, \citenamefont {Ma}, \citenamefont {Schmidt}, \citenamefont {Greiter}, \citenamefont {Kiessling}, \citenamefont {Szameit},\ and\ \citenamefont {Thomale}}]{Stegmaier2021}%
  \BibitemOpen
  \bibfield  {author} {\bibinfo {author} {\bibfnamefont {A.}~\bibnamefont {Stegmaier}}, \bibinfo {author} {\bibfnamefont {S.}~\bibnamefont {Imhof}}, \bibinfo {author} {\bibfnamefont {T.}~\bibnamefont {Helbig}}, \bibinfo {author} {\bibfnamefont {T.}~\bibnamefont {Hofmann}}, \bibinfo {author} {\bibfnamefont {C.~H.}\ \bibnamefont {Lee}}, \bibinfo {author} {\bibfnamefont {M.}~\bibnamefont {Kremer}}, \bibinfo {author} {\bibfnamefont {A.}~\bibnamefont {Fritzsche}}, \bibinfo {author} {\bibfnamefont {T.}~\bibnamefont {Feichtner}}, \bibinfo {author} {\bibfnamefont {S.}~\bibnamefont {Klembt}}, \bibinfo {author} {\bibfnamefont {S.}~\bibnamefont {H\"ofling}}, \bibinfo {author} {\bibfnamefont {I.}~\bibnamefont {Boettcher}}, \bibinfo {author} {\bibfnamefont {I.~C.}\ \bibnamefont {Fulga}}, \bibinfo {author} {\bibfnamefont {L.}~\bibnamefont {Ma}}, \bibinfo {author} {\bibfnamefont {O.~G.}\ \bibnamefont {Schmidt}}, \bibinfo {author} {\bibfnamefont {M.}~\bibnamefont {Greiter}}, \bibinfo {author} {\bibfnamefont {T.}~\bibnamefont
  {Kiessling}}, \bibinfo {author} {\bibfnamefont {A.}~\bibnamefont {Szameit}}, \ and\ \bibinfo {author} {\bibfnamefont {R.}~\bibnamefont {Thomale}},\ }\href {\doibase 10.1103/PhysRevLett.126.215302} {\bibfield  {journal} {\bibinfo  {journal} {Phys. Rev. Lett.}\ }\textbf {\bibinfo {volume} {126}},\ \bibinfo {pages} {215302} (\bibinfo {year} {2021})}\BibitemShut {NoStop}%
\bibitem [{\citenamefont {Zou}\ \emph {et~al.}(2021)\citenamefont {Zou}, \citenamefont {Chen}, \citenamefont {He}, \citenamefont {Bao}, \citenamefont {Lee}, \citenamefont {Sun},\ and\ \citenamefont {Zhang}}]{Zou2021}%
  \BibitemOpen
  \bibfield  {author} {\bibinfo {author} {\bibfnamefont {D.}~\bibnamefont {Zou}}, \bibinfo {author} {\bibfnamefont {T.}~\bibnamefont {Chen}}, \bibinfo {author} {\bibfnamefont {W.}~\bibnamefont {He}}, \bibinfo {author} {\bibfnamefont {J.}~\bibnamefont {Bao}}, \bibinfo {author} {\bibfnamefont {C.~H.}\ \bibnamefont {Lee}}, \bibinfo {author} {\bibfnamefont {H.}~\bibnamefont {Sun}}, \ and\ \bibinfo {author} {\bibfnamefont {X.}~\bibnamefont {Zhang}},\ }\href {\doibase 10.1038/s41467-021-26414-5} {\bibfield  {journal} {\bibinfo  {journal} {Nat. Commun.}\ }\textbf {\bibinfo {volume} {12}},\ \bibinfo {pages} {7201} (\bibinfo {year} {2021})}\BibitemShut {NoStop}%
\bibitem [{\citenamefont {Yatsugi}\ \emph {et~al.}(2022)\citenamefont {Yatsugi}, \citenamefont {Yoshida}, \citenamefont {Mizoguchi}, \citenamefont {Kuno}, \citenamefont {Iizuka}, \citenamefont {Tadokoro},\ and\ \citenamefont {Hatsugai}}]{Yatsugi2022}%
  \BibitemOpen
  \bibfield  {author} {\bibinfo {author} {\bibfnamefont {K.}~\bibnamefont {Yatsugi}}, \bibinfo {author} {\bibfnamefont {T.}~\bibnamefont {Yoshida}}, \bibinfo {author} {\bibfnamefont {T.}~\bibnamefont {Mizoguchi}}, \bibinfo {author} {\bibfnamefont {Y.}~\bibnamefont {Kuno}}, \bibinfo {author} {\bibfnamefont {H.}~\bibnamefont {Iizuka}}, \bibinfo {author} {\bibfnamefont {Y.}~\bibnamefont {Tadokoro}}, \ and\ \bibinfo {author} {\bibfnamefont {Y.}~\bibnamefont {Hatsugai}},\ }\href {\doibase 10.1038/s42005-022-00957-5} {\bibfield  {journal} {\bibinfo  {journal} {Comm. Phys.}\ }\textbf {\bibinfo {volume} {5}},\ \bibinfo {pages} {180} (\bibinfo {year} {2022})}\BibitemShut {NoStop}%
\bibitem [{\citenamefont {Kotwal}\ \emph {et~al.}(2021)\citenamefont {Kotwal}, \citenamefont {Moseley}, \citenamefont {Stegmaier}, \citenamefont {Imhof}, \citenamefont {Brand}, \citenamefont {Kießling}, \citenamefont {Thomale}, \citenamefont {Ronellenfitsch},\ and\ \citenamefont {Dunkel}}]{Kotwal2021}%
  \BibitemOpen
  \bibfield  {author} {\bibinfo {author} {\bibfnamefont {T.}~\bibnamefont {Kotwal}}, \bibinfo {author} {\bibfnamefont {F.}~\bibnamefont {Moseley}}, \bibinfo {author} {\bibfnamefont {A.}~\bibnamefont {Stegmaier}}, \bibinfo {author} {\bibfnamefont {S.}~\bibnamefont {Imhof}}, \bibinfo {author} {\bibfnamefont {H.}~\bibnamefont {Brand}}, \bibinfo {author} {\bibfnamefont {T.}~\bibnamefont {Kießling}}, \bibinfo {author} {\bibfnamefont {R.}~\bibnamefont {Thomale}}, \bibinfo {author} {\bibfnamefont {H.}~\bibnamefont {Ronellenfitsch}}, \ and\ \bibinfo {author} {\bibfnamefont {J.}~\bibnamefont {Dunkel}},\ }\href {\doibase 10.1073/pnas.2106411118} {\bibfield  {journal} {\bibinfo  {journal} {Proc. Nat. Aca. Sci.}\ }\textbf {\bibinfo {volume} {118}},\ \bibinfo {pages} {e2106411118} (\bibinfo {year} {2021})}\BibitemShut {NoStop}%
\bibitem [{\citenamefont {Stegmaier}\ \emph {et~al.}(2024)\citenamefont {Stegmaier}, \citenamefont {Brand}, \citenamefont {Imhof}, \citenamefont {Fritzsche}, \citenamefont {Helbig}, \citenamefont {Hofmann}, \citenamefont {Boettcher}, \citenamefont {Greiter}, \citenamefont {Lee}, \citenamefont {Bahl}, \citenamefont {Szameit}, \citenamefont {Kie\ss{}ling}, \citenamefont {Thomale},\ and\ \citenamefont {Upreti}}]{stegmaier2023thouless}%
  \BibitemOpen
  \bibfield  {author} {\bibinfo {author} {\bibfnamefont {A.}~\bibnamefont {Stegmaier}}, \bibinfo {author} {\bibfnamefont {H.}~\bibnamefont {Brand}}, \bibinfo {author} {\bibfnamefont {S.}~\bibnamefont {Imhof}}, \bibinfo {author} {\bibfnamefont {A.}~\bibnamefont {Fritzsche}}, \bibinfo {author} {\bibfnamefont {T.}~\bibnamefont {Helbig}}, \bibinfo {author} {\bibfnamefont {T.}~\bibnamefont {Hofmann}}, \bibinfo {author} {\bibfnamefont {I.}~\bibnamefont {Boettcher}}, \bibinfo {author} {\bibfnamefont {M.}~\bibnamefont {Greiter}}, \bibinfo {author} {\bibfnamefont {C.~H.}\ \bibnamefont {Lee}}, \bibinfo {author} {\bibfnamefont {G.}~\bibnamefont {Bahl}}, \bibinfo {author} {\bibfnamefont {A.}~\bibnamefont {Szameit}}, \bibinfo {author} {\bibfnamefont {T.}~\bibnamefont {Kie\ss{}ling}}, \bibinfo {author} {\bibfnamefont {R.}~\bibnamefont {Thomale}}, \ and\ \bibinfo {author} {\bibfnamefont {L.~K.}\ \bibnamefont {Upreti}},\ }\href {\doibase 10.1103/PhysRevResearch.6.023010} {\bibfield  {journal} {\bibinfo  {journal} {Phys. Rev.
  Res.}\ }\textbf {\bibinfo {volume} {6}},\ \bibinfo {pages} {023010} (\bibinfo {year} {2024})}\BibitemShut {NoStop}%
\bibitem [{app()}]{app}%
  \BibitemOpen
  \bibinfo {note} {{Appendix containing a detailed derivation of all circuit dynamics and a complete listing of the explicit components used to build the Floquet SSH circuit}}\BibitemShut {NoStop}%
\bibitem [{\citenamefont {Oka}\ and\ \citenamefont {Kitamura}(2019)}]{Oka2019engineerQM}%
  \BibitemOpen
  \bibfield  {author} {\bibinfo {author} {\bibfnamefont {T.}~\bibnamefont {Oka}}\ and\ \bibinfo {author} {\bibfnamefont {S.}~\bibnamefont {Kitamura}},\ }\href {\doibase 10.1146/annurev-conmatphys-031218-013423} {\bibfield  {journal} {\bibinfo  {journal} {Annu. Rev. Condens. Matter Phys.}\ }\textbf {\bibinfo {volume} {10}},\ \bibinfo {pages} {387} (\bibinfo {year} {2019})}\BibitemShut {NoStop}%
\bibitem [{\citenamefont {Qiao}\ \emph {et~al.}(2023)\citenamefont {Qiao}, \citenamefont {Wang}, \citenamefont {Li}, \citenamefont {Chen},\ and\ \citenamefont {Yuan}}]{Qiao2023degeneracy}%
  \BibitemOpen
  \bibfield  {author} {\bibinfo {author} {\bibfnamefont {X.}~\bibnamefont {Qiao}}, \bibinfo {author} {\bibfnamefont {L.}~\bibnamefont {Wang}}, \bibinfo {author} {\bibfnamefont {G.}~\bibnamefont {Li}}, \bibinfo {author} {\bibfnamefont {X.}~\bibnamefont {Chen}}, \ and\ \bibinfo {author} {\bibfnamefont {L.}~\bibnamefont {Yuan}},\ }\href {\doibase doi:10.1515/nanoph-2023-0408} {\bibfield  {journal} {\bibinfo  {journal} {Nanophotonics}\ }\textbf {\bibinfo {volume} {12}},\ \bibinfo {pages} {3807} (\bibinfo {year} {2023})}\BibitemShut {NoStop}%
\bibitem [{\citenamefont {Li}\ \emph {et~al.}(2023)\citenamefont {Li}, \citenamefont {Wang}, \citenamefont {Ye}, \citenamefont {Zheng}, \citenamefont {Wang}, \citenamefont {Liu}, \citenamefont {Dutt}, \citenamefont {Yuan},\ and\ \citenamefont {Chen}}]{li2023zak}%
  \BibitemOpen
  \bibfield  {author} {\bibinfo {author} {\bibfnamefont {G.}~\bibnamefont {Li}}, \bibinfo {author} {\bibfnamefont {L.}~\bibnamefont {Wang}}, \bibinfo {author} {\bibfnamefont {R.}~\bibnamefont {Ye}}, \bibinfo {author} {\bibfnamefont {Y.}~\bibnamefont {Zheng}}, \bibinfo {author} {\bibfnamefont {D.-W.}\ \bibnamefont {Wang}}, \bibinfo {author} {\bibfnamefont {X.-J.}\ \bibnamefont {Liu}}, \bibinfo {author} {\bibfnamefont {A.}~\bibnamefont {Dutt}}, \bibinfo {author} {\bibfnamefont {L.}~\bibnamefont {Yuan}}, \ and\ \bibinfo {author} {\bibfnamefont {X.}~\bibnamefont {Chen}},\ }\href {\doibase 10.1038/s41377-023-01126-1} {\bibfield  {journal} {\bibinfo  {journal} {Light Sci. Appl.}\ }\textbf {\bibinfo {volume} {12}},\ \bibinfo {pages} {81} (\bibinfo {year} {2023})}\BibitemShut {NoStop}%
\bibitem [{\citenamefont {Song}\ \emph {et~al.}(2023)\citenamefont {Song}, \citenamefont {Ke}, \citenamefont {Chen},\ and\ \citenamefont {Wang}}]{Song2023AntiPTSSH}%
  \BibitemOpen
  \bibfield  {author} {\bibinfo {author} {\bibfnamefont {Y.}~\bibnamefont {Song}}, \bibinfo {author} {\bibfnamefont {S.}~\bibnamefont {Ke}}, \bibinfo {author} {\bibfnamefont {Y.}~\bibnamefont {Chen}}, \ and\ \bibinfo {author} {\bibfnamefont {M.}~\bibnamefont {Wang}},\ }\href {\doibase 10.1063/5.0146246} {\bibfield  {journal} {\bibinfo  {journal} {Appl. Phys. Lett.}\ }\textbf {\bibinfo {volume} {122}},\ \bibinfo {pages} {151106} (\bibinfo {year} {2023})}\BibitemShut {NoStop}%
\end{thebibliography}
\end{document}